\documentclass{ieeeaccess}
\usepackage{cite}
\usepackage{amsmath,amssymb,amsfonts}
\usepackage{algorithmic}
\usepackage{graphicx}
\usepackage{textcomp}

\usepackage{graphics} % for pdf, bitmapped graphics files
\usepackage{epsfig} % for postscript graphics files
\usepackage{mathptmx} % assumes new font selection scheme installed
\usepackage{times} % assumes new font selection scheme installed
\usepackage{amsmath} % assumes amsmath package installed
\usepackage{amssymb}  % assumes amsmath package installed
\usepackage{color}
\usepackage{multirow}
\usepackage{footnote}
\usepackage{caption}
\usepackage[hyphens]{url} % Marc
\usepackage{hyperref}
\usepackage{float}
\usepackage{cuted}
\usepackage[table,xcdraw]{xcolor}
\usepackage{subcaption}
\usepackage{comment}
\usepackage{cite}
\usepackage[utf8]{inputenc} % accents

\def\BibTeX{{\rm B\kern-.05em{\sc i\kern-.025em b}\kern-.08em
    T\kern-.1667em\lower.7ex\hbox{E}\kern-.125emX}}
\begin{document}
\history{Received January 29, 2021, accepted February 11, 2021. Date of publication xxxx 00, 0000, date of current version xxxx 00, 0000.}
\doi{10.1109/ACCESS.2021.3059473}

\title{Channel load aware AP / Extender selection in Home WiFi networks using IEEE 802.11k/v}
\author{\uppercase{Toni Adame}\authorrefmark{1},
\uppercase{Marc Carrascosa\authorrefmark{1}, Boris Bellalta\authorrefmark{1}, Iv\'{a}n Pretel\authorrefmark{2}, and I\~{n}aki Etxebarria}.\authorrefmark{2}}
\address[1]{Department of Information and Communication Technologies, Universitat Pompeu Fabra. Carrer de Roc Boronat 138, 08018 Barcelona, Spain.}
\address[2]{FON Labs. Plaza Euskadi 5, P15-15, 48009 Bilbao, Spain.}
\tfootnote{This work received funding from the Spanish government under the projects CDTI IDI-20180274, WINDMAL PGC2018-099959-B-100 (MCIU/AEI/FEDER,UE), TEC2016-79510-P, and from the Catalan government through projects SGR-2017-1188 and SGR-2017-1739.}

\markboth
{T. Adame \headeretal: Channel load aware AP / Extender selection in Home WiFi networks using IEEE 802.11k/v}
{T. Adame \headeretal: Channel load aware AP / Extender selection in Home WiFi networks using IEEE 802.11k/v}

\corresp{Corresponding author: Toni Adame (e-mail: toni.adame@upf.edu).}

\begin{abstract}
Next-generation Home WiFi networks have to step forward in terms of performance. New applications such as on-line games, virtual reality or high quality video contents will further demand higher throughput levels, as well as low latency. Beyond physical (PHY) and medium access control (MAC) improvements, deploying multiple access points (APs) in a given area may significantly contribute to achieve those performance goals by simply improving average coverage and data rates. However, it opens a new challenge: to determine the best AP for each given station (STA). This article studies the achievable performance gains of using secondary APs, also called Extenders, in Home WiFi networks in terms of throughput and delay. To do that, we introduce a centralized, easily implementable \textit{channel load aware} selection mechanism for WiFi networks that takes full advantage of IEEE 802.11k/v capabilities to collect data from STAs, and distribute association decisions accordingly. These decisions are completely computed in the AP (or, alternatively, in an external network controller) based on an AP selection decision metric that, in addition to RSSI, also takes into account the load of both access and backhaul wireless links for each potential STA-AP/Extender connection. Performance evaluation of the proposed \textit{channel load aware} AP and Extender selection mechanism has been first conducted in a purpose-built simulator, resulting in an overall improvement of the main analyzed metrics (throughput and delay) and the ability to serve, at least, 35\% more traffic while keeping the network uncongested when compared to the traditional \textit{RSSI-based} WiFi association. This trend was confirmed when the \textit{channel load aware} mechanism was tested in a real deployment, where STAs were associated to the indicated AP/Extender and total throughput was increased by 77.12\%.
\end{abstract}


\titlepgskip=-15pt

\maketitle

\section{Introduction}
\label{sec:introduction}

% Intro per explicar que és una WIFI i motivar l'interés en fer servir extenders

\PARstart{S}{ince} their appearance more than 20 years ago, IEEE 802.11 wireless local area networks (WLANs) have become the worldwide preferred option to provide wireless Internet access to heterogeneous clients in homes, businesses, and public spaces due to their low cost and mobility support. The simplest WLAN contains only a basic service set (BSS), consisting of an access point (AP) connected to a wired infrastructure, and some wireless stations (STAs) associated to the AP. 

% Future challenges: more throughput

The increase of devices aiming to use the WLAN technology to access Internet has been accompanied by more demanding user requirements, especially in entertainment contents: on-line games, virtual reality, and high quality video. In consequence, traditional single-AP WLANs deployed in apartments, i.e., Home WiFi networks, may fail to deliver a satisfactory experience due to the existence of areas where the received power from the AP is low, and so the achievable performance \cite{hoiland2017ending}.

Although IEEE 802.11ac (WiFi 5) \cite{6687187}, IEEE 802.11ax (WiFi 6) \cite{bellalta2016ieee,khorov2018tutorial}, and IEEE 802.11be (WiFi 7) \cite{lopez2019ieee,adame2019time} amendments provide enhancements on physical (PHY) and medium access control (MAC) protocols that may increase the WLAN efficiency, and also increase the coverage by using beamforming, the best solution is still to deploy more APs to improve the coverage in those areas.

% Wireless Extenders

In multi-AP deployments, normally only one AP (the \textit{main} AP) has Internet access, and so the other APs (from now on simply called \textit{Extenders}) must relay the data to it using a wired or wireless backhaul network. Since presuming the existence of a wired network is not always feasible, Extenders communicate with the main AP wirelessly. In this case, both the main AP and Extenders are equipped with at least two radios, usually operating at different bands.

% L'elecció de l'AP o del extender per associar-te

In presence of multiple AP/Extenders, a new challenge appears: how to determine the best AP/Extender for each given STA. According to the default WiFi AP selection mechanism, an STA that receives beacons from several AP/Extenders will initiate the association process with the AP/Extender with the highest received signal strength indicator (RSSI) value. Though simple and easy to implement, this mechanism omits any influence of traffic load and, consequently, can lead to network congestion and low throughput in scenarios with a high number of STAs \cite{yen2009load}.

Many research activities have already widely tackled the AP selection process in an area commonly referred to as \textit{load balancing}, whose goal is to distribute more efficiently STAs among the available AP/Extenders in a WLAN. Although multiple effective strategies have been proposed in the literature, most of them lack the prospect of real implementation, as they require changes in the existing IEEE 802.11 standards and/or in STAs' wireless cards. 

The \textit{channel load aware} AP/Extender selection mechanism presented in this article sets out to enhance the overall WLAN performance by including the effect of the channel load into the STA association process. To do it, only already developed IEEE 802.11 amendments are considered: IEEE 802.11k to gather information from AP/Extenders in the WLAN, and IEEE 802.11v to notify each STA of its own prioritized list of AP/Extenders. 

% Les nostres contribucions

Particularly, the main contributions of the current work can be summarized into:

\begin{itemize}
    \item Review and classification of multiple existing AP/Extender selection mechanisms, and some background information on the use of IEEE 802.11k/v.
    \item Design of a feasible, practical, and flexible \textit{channel load aware} AP/Extender selection mechanism supported by IEEE 802.11k/v amendments.
    \item Evaluation of the \textit{channel load aware} AP/Extender selection mechanism by simulation, studying the performance gains of using Extenders along with the proposed solution. We focus on understanding how the number of Extenders and their position, the fraction of STAs supporting IEEE 802.11k/v, and the load of the access and backhaul links, impact on the system performance in terms of throughput and delay.
    \item Validation of the presented solution in a real testbed, showing the same trends in terms of performance improvements that those obtained by simulation.
\end{itemize}

Lastly, the main lessons that can be learned from this article are listed below:
\begin{enumerate}
    \item Placement of Extenders: We observe that Extenders must be located at a distance (in RSSI terms) large enough to stimulate the association of farther STAs while maintaining high data rate in its backhaul connection to the AP. Also, we confirm that connecting Extenders through other Extenders not only increases the network coverage, but also the network's operational range in terms of admitted traffic load.
    \item Load of access vs. backhaul links: The relative weight of the load of the access and backhaul(s) link(s) should be generally balanced, without dismissing a proper tuning according to the characteristics of the deploying scenario.
    \item STAs supporting IEEE 802.11k/v: We observe that, even for a low fraction of STAs supporting IEEE 802.11k/v, the gains of using the \textit{channel load aware} AP/Extender selection mechanism are beneficial for the overall network.
    \item Throughput and delay improvements: The use of Extenders allows to balance the load of the network, which results in significant gains in throughput and delay for much higher traffic loads. Therefore, the use of Extenders is recommended for high throughput multimedia and delay-constrained applications. 
\end{enumerate}

% L'estructura del paper

The remainder of this article is organized as follows: Section~\ref{selection} offers an overview on AP selection in WiFi networks. Section~\ref{11kv} elaborates on IEEE 802.11k and IEEE 802.11v amendments, paying special attention to the features considered in the proposed \textit{channel load aware} AP/Extender selection mechanism, which is in turn described in Section \ref{load_aware}. Performance results obtained from simulations and real deployments are compiled in Section \ref{simulation} and Section \ref{deployment}, respectively. Lastly, Section \ref{challenges} discusses open challenges in future Home WiFi networks and Section \ref{conclusions} presents the obtained conclusions.

%----------------------------------------------------------------------------------
%----------------------------------------------------------------------------------
%----------------------------------------------------------------------------------
%----------------------------------------------------------------------------------

\section{Use of Extenders and AP/Extender selection mechanisms in WiFi networks}
\label{selection}

The current section reviews the main aspects of the technical framework involving the use of Extenders in next-generation Home WiFi networks, such as the main challenges related to their deployment, the existing options to integrate them into the STA association procedure, and their management through an external platform.

% ------------------------------
% ------------------------------

\subsection{Multi-hop communication in WLANs}

The need to expand WLAN coverage to every corner of a targeted area can be satisfied by increasing the AP transmission power or by deploying wired/wireless Extenders. Putting aside the wired option, which is not in scope of the current article, wireless extension of a WLAN can be achieved by means of a wireless mesh network (WMN).

In a WMN, multiple deployed APs communicate among them in a multi-hop scheme to relay data from/to STAs. The most representative initiative in this field is IEEE 802.11s, which integrates mesh networking services and protocols with IEEE 802.11 at the MAC layer \cite{hiertz2010ieee}. Wireless frame forwarding and routing capabilities are managed by the hybrid wireless mesh protocol (HWMP), which combines the flexibility of on-demand route discovery with efficient proactive routing to a mesh portal \cite{bari2012performance}.

As traffic streams in a WMN are mainly oriented towards/from the main AP, they tend to form a tree-based wireless architecture \cite{waharte2005tree}. This architecture strongly relies on the \textit{optimal} number and position of deployed Extenders, which is determined in \cite{herlich2014optimal} as a function of PHY layer parameters with the goal of minimizing latency and maximizing data rate. This analysis is extended in \cite{hassan2013optimal}, where a model based on PHY and MAC parameters returns those Extender locations that maximize multi-hop throughput. Other approaches such as \cite{atawia2017self} go far beyond and propose the use of Artifical Intelligence to enable autonomous self-deployment of wireless Extenders.

Relaying capabilities of Extenders are also a matter of study, as in \cite{fadlullah2017multi}, where an algorithm is proposed to determine the optimal coding rate and modulation scheme to dynamically control the best band and channel selection. Or in  \cite{egashira2017low}, where a low latency relay transmission scheme for WLAN is proposed to  simultaneously use multiple frequency bands.

All in all, once the number, location and relaying capabilities of Extenders operating in a WLAN are selected, the way in which STAs determine their own parent (i.e., the best AP/Extender located within their coverage area) can impact the overall performance of the network. We will discuss on this issue in the following lines.

% ------------------------------
% ------------------------------

\subsection{AP/Extender selection mechanisms}

A review of the currently existing AP/Extender selection mechanisms along with the description of the WiFi scanning modes that enable them is offered in the following lines.

\subsubsection{WiFi scanning modes}

IEEE 802.11 standard defines two different scanning modes: \textit{passive} and \textit{active} \cite{7786995}. In \textit{passive scanning}, for each available radio channel, the STA listens to beacons sent by APs for a dwell time. As beacons are usually broadcast by the AP every 100 ms, channel dwell time is typically set to 100-200 ms to guarantee beacon reception \cite{chen2010hand,choi2017energy}.

In \textit{active scanning}, the STA starts broadcasting a \textit{probe request} frame on one channel and sets a probe timer. If no \textit{probe response} is received before the probe timer reaches \textit{MinChannelTime}, the STA assumes that no AP is working in that channel and scans another channel alternatively. Otherwise, if the STA does receive a \textit{probe response}, it will further wait for responses from other working APs until \textit{MaxChannelTime} is reached by the probe timer. \textit{MinChannelTime} and \textit{MaxChannelTime} values are vendor-specific, as they are not specified by the IEEE 802.11 standard. Indeed, the obtention of optimum values to minimize the \textit{active scanning} phase have attracted research attention. In \cite{mishra2003empirical}, for instance, the author sets these values as low as 6-7 ms and 10-15 ms, respectively.

Since \textit{passive scanning} always has longer latency than \textit{active scanning}, wireless cards tend to use the latter to rapidly find nearby APs \cite{velayos2004techniques}. However, \textit{active scanning} has three disadvantages: 1) it consumes significant more energy than \textit{passive scanning}, 2) it is unable to discover networks that do not broadcast their SSID, and 3) it may result in shorter scan ranges because of the lower power level of STAs.

It is also usual that mobile STAs periodically perform \textit{active background scanning} to discover available APs, and then accelerate an eventual roaming operation \cite{lee2012smart}. In this case, the STA (already associated to an AP and exchanging data) goes periodically off-channel and sends probe requests across other channels. On the other hand, the \textit{active on-roam scanning} only occurs after the STA determines a roam is necessary.

%MORE INFO ON THIS:
%https://www.cisco.com/c/en/us/td/docs/solutions/Enterprise/Mobility/RToWLAN/CCVP_BK_R7805F20_00_rtowlan-srnd/CCVP_BK_R7805F20_00_rtowlan-srnd_chapter_0101.pdf

%During a channel scan, the client is unable to transmit or receive client data traffic. Clients use the following approaches to minimize this impact to client data traffic:
%• Background scanning: Clients scan the available channels before they roam. The scans provide information about the RF environment and available APs that can help clients to roam faster, if necessary. The affect to client traffic can be minimized by scanning only when the client is not actively transmitting data, or by periodically scanning only a single alternate channel at a time. Scanning a single channel incurs minimal data loss.

%• On-roam scanning: On-roam scan occurs after the client determines a roam is necessary. Each vendor or device can implement its own algorithms to minimize the roam latency and the affect to data traffic. For example, some clients might scan only the non-overlapping channels.

\subsubsection{Default WiFi AP selection mechanism}

Regardless the scanning mode used by an STA to complete its own list of available APs, and the final purpose of this scanning (i.e., the initial association after the STA startup or a roaming operation), the STA executes the default WiFi AP selection mechanism (from now on also named \textit{RSSI-based}) by choosing the AP of the previous list with the strongest RSSI.

This is the approach followed by common APs and available multi-AP commercial solutions, like Google WiFi \cite{google2019wifi} or Linksys Velop \cite{linksys2019velop}, which are especially indicated for homes with coverage problems and few users. In addition, these two solutions also integrate the IEEE 802.11k/v amendments (analyzed later on in Section \ref{11kv}), but only to provide faster and seamless roaming.

%REF: https://www.smallnetbuilder.com/wireless/wireless-features/33207-wi-fi-roaming-secrets-revealed-part-4?start=1

The strongest RSSI might indicate the best channel condition between the STA and the AP. However, only relying on this criteria is not always the best choice, as it can lead to imbalanced loads between APs, inefficient rate selection, and selection of APs with poor throughput, delay, and other performance metrics \cite{ernst2014utility}.

\begin{table}[t]
\footnotesize
\centering
\caption{Classification of alternative AP/Extender selection mechanisms. Whereas mechanisms employing to some extent IEEE 802.11k are marked with \dag, no mechanism employs IEEE 802.11v. (By default, parameters from the decision metric column refer to the STA's value).}
\label{tab:classification}
\begin{tabular}{|c|c|c|c|}
\hline
\multirow{2}{*}{\textbf{Mechanism}} & \multicolumn{3}{c|}{\textbf{Classification criteria}}                                                                         \\ \cline{2-4} 
                                    & \begin{tabular}[c]{@{}c@{}}\textbf{AP selection} \\ \textbf{mode}\end{tabular} & \textbf{Architecture} & \textbf{Decision metric}                        \\ \hline
                       \cite{nicholson2006improved}          & \textit{Active}                                                     & \textit{Decentralized}                                                  & \begin{tabular}[c]{@{}c@{}}Bandwidth, RTT, and\\ available ports\end{tabular}                                  \\ \hline
                  
                  \cite{yen2011stability}       &   \textit{Active}                                           &  \textit{Decentralized}                                             & Throughput  \\ \hline
\cite{fukuda2007decentralized}               & \textit{Active}                                               & \textit{Decentralized}                                             &  Throughput or PER 
                       
                               \\ \hline
  \cite{gong2008distributed}        & \textit{Active}                                            & \textit{Decentralized}                                                 & AP load                              

  \\ \hline
                                    \cite{xu2010designing}
                                    & \textit{Active}                                                    & \textit{Decentralized}                                                  & AP load
   \\ \hline
                        \cite{raschella2016centralized}            & \textit{Active}                                                     & \textit{Centralized}                                                  & \begin{tabular}[c]{@{}c@{}}Fittingness factor\\ (mainly based on rate)\end{tabular}    
  
  \\ \hline
                             \cite{abusubaih2007optimal} \dag       & \textit{Active}                                              &                  \textit{Centralized}                            &       WLAN throughput                           \\ \hline
         \cite{du2007access}       &   
         \begin{tabular}[c]{@{}c@{}}\textit{Active} or\\ \textit{passive}\end{tabular}  
         
         \                                           & \textit{Decentralized}                                              &                       \begin{tabular}[c]{@{}c@{}}Contention with\\ hidden terminals\end{tabular}                                              \\ \hline
                    \cite{abusubaih2006access} \dag      &    \begin{tabular}[c]{@{}c@{}}\textit{Active} or\\ \textit{passive}\end{tabular}                                           &   \textit{Decentralized}                                        &    \begin{tabular}[c]{@{}c@{}}Throughput and channel\\ occupancy rate\end{tabular}
                       
                       \\ \hline
                           
                                         \cite{ernst2014utility} &  \textit{Passive}      &               \textit{Decentralized}                               &
                       \begin{tabular}[c]{@{}c@{}}Distance, rate, delay,\\ or a combination of them \end{tabular}                                                          \\ \hline
                          \cite{vasudevan2005facilitating}          & \textit{Passive}                                                    & \textit{Decentralized}                                                  & Bandwidth    
                          \\ \hline
                   \cite{mittal2008game}       & \textit{Passive}                                              &                  \textit{Decentralized}                            &    
                  Distance and AP load   
                                    \\ \hline
                  
                                    \cite{luo2008improving}       &   \textit{Passive}                                           &  \textit{Decentralized}                                             &                 Transmission time
                                    \\ \hline
                    \cite{abusubaih2008interference} \dag      &    \textit{Hybrid}                                         &   \textit{Decentralized}                                        &   \begin{tabular}[c]{@{}c@{}}Throughput and \\ throughput impact \\ on other STAs\end{tabular}
                    
                    \\ \hline

       \cite{sundaresan2006need}       & \textit{Hybrid}                                              &                  \textit{Decentralized}                            &       Throughput        \\ \hline
                                \cite{pang2010wifi}       & \textit{Hybrid}                                              &                  \textit{Centralized}                            &    
                  \begin{tabular}[c]{@{}c@{}}Throughput, delay,\\ and connection state\end{tabular}           
            
                  \\ \hline
\end{tabular}
\end{table}

\subsubsection{Alternative AP/Extender selection mechanisms}

The inefficiency of the \textit{RSSI-based} AP selection mechanism has motivated the emergence of alternative methods that take into account other metrics than solely the RSSI. The most representative examples are compiled in Table \ref{tab:classification} and classified according to three different criteria: the AP selection mode, the architecture employed, and the selected decision metric:

\begin{itemize}
    \item \textbf{AP selection mode}: In the \textit{active} AP selection, the STA considers all potential APs and gathers information regarding one or more performance metrics to make a decision. In \cite{nicholson2006improved}, the STA scans for all available APs, quickly associates to each, and even runs a set of tests to estimate Internet connection quality. On the contrary, the \textit{passive} AP selection is based on the information that the STA directly extracts from beacon frames or deduces from their physical features, such as the experienced delay in \cite{vasudevan2005facilitating}. \\Lastly, in the \textit{hybrid} AP selection, the network makes use of the information shared by the STA to give advice on the best possible potential AP. In \cite{pang2010wifi}, for instance, clients automatically submit reports on the APs that they use with regard to estimated backhaul capacity, ports blocked, and connectivity failures.
    \item \textbf{Architecture}: This category splits the different mechanisms into \textit{decentralized} and \textit{centralized}. \textit{Decentralized} mechanisms are those in which the STA selects its AP based on its available information (even combining cross-layer information, as in \cite{sundaresan2006need}). On the other hand, \textit{centralized} mechanisms imply a certain degree of coordination between different APs thanks to a central entity (that may well be an SDN controller, as in \cite{raschella2016centralized}) intended to balance overall network load.
    \item \textbf{Decision metric}: The AP selection metric can be determined by a single parameter (e.g., AP load in \cite{gong2008distributed}) or a weighted combination of some of them (e.g., throughput and channel occupancy rate in \cite{abusubaih2006access}). Apart from RSSI, there exists a vast quantity of available magnitudes for this purpose; however, the most common ones in the reviewed literature are throughput, load, and delay.
\end{itemize}

Furthermore, there exist some novel approaches that have introduced machine learning (ML) techniques into the AP selection process. For instance, in \cite{bojovic2012neural} a \textit{decentralized} cognitive engine based on a neural network trained on past link conditions and throughput performance drives the AP selection process. 

Likewise, a decentralized approach based on the exploration-exploitation trade-off from Reinforcement Learning algorithms is used in \cite{8969724,carrascosa2020multi}. Under that system, STAs learn the network conditions and associate to the AP that maximizes their throughput. In consequence, STAs stop its exploration, which is only resumed when there is a change in network's topology. 

Another \textit{decentralized} ML-based approach is proposed in \cite{chen2010distributed}, where the AP selection mechanism is formulated as a non-cooperative game in which each STA tries to maximize its throughput. Then, an adaptive algorithm based on no-regret learning makes the system converge to an equilibrium state.

% ------------------------------
% ------------------------------

\subsection{Commercial WLAN Management Platforms}

Centralized network management platforms are commonly used in commercial solutions, as they give full control of the network to the operator. These management platforms focus not only on the AP selection, but also cover several network performance enhancements such as channel and band selection, and transmit power adjustment.

Nighthawk Mesh WiFi 6 System \cite{netgear2019mk62} intelligently selects the fastest WiFi band for every connected STA, and Insight Management Solution \cite{netgear2018rrm} recalculates the optimum channel and transmit power for all the APs every 24 hours. %In addition, another platform relies on Artificial Intelligence algorithms to automatically optimize radio and logic resources. 
Based on signal strength and channel utilization metrics, ArubaOS network operating system has components (i.e. AirMatch \cite{aruba2018airmatch} and ClientMatch \cite{aruba2018clientmatch}) which dynamically balance STAs across channels and encourage dual-band capable STAs to stay on the 5GHz band on dual-band APs. Lastly, Cognitive Hotspot Technology (CHT) \cite{garrido2019system} is a multi-platform software that can be installed on a wide range of APs. It brings distributed intelligence to any WiFi network to control the radio resources including AP automatic channel selection, load balancing, as well as client and band steering for STAs.

The \textit{channel load aware} AP/Extender selection mechanism presented in this work could be easily integrated in these centralized platforms and even be further enhanced by exploiting the \textit{know-how} gathered from different Home WiFi networks.

%----------------------------------------------------------------------------------
%----------------------------------------------------------------------------------
%----------------------------------------------------------------------------------
%----------------------------------------------------------------------------------

\section{IEEE 802.11k/v amendments}
\label{11kv}

The constant evolution of the IEEE 802.11 standard has been fostered by the incremental incorporation of technical amendments addressing different challenges in the context of WLANs. In particular, the optimization of the AP selection process and the minimization of the roaming interruption time are tackled in two different amendments: IEEE 802.11k and IEEE 802.11v \cite{sanchez2016ieee}.

\subsection{IEEE 802.11k: Radio Resource Measurement}

The IEEE 802.11k amendment on radio resource measurement \cite{4544755} defines methods for information exchange about the radio environment between APs and STAs. This information may be thus used for radio resource management strategies, making devices more likely to properly adapt to the dynamic radio environment.  

Radio environment information exchange between two devices running IEEE 802.11k occurs through a two-part frame request/report exchange carried within radio measurement report frames (i.e., a purpose-specific category of action frames). Despite the wide set of possible measurements, the AP/Extender selection mechanism presented in this work will only consider \textit{beacon} and \textit{channel load reports}.

The \textit{beacon request/report} pair enables an AP to ask an STA for the list of APs it is able to listen effectively to on a specified channel or channels. The request also includes the measurement mode that should be performed by the targeted STA: \textit{active scanning} (i.e., information comes from \textit{probe responses}), \textit{passive scanning} (i.e., information comes from \textit{beacons}), or \textit{beacon table} (i.e., use of previously stored beacon information).

Whenever an STA receives a \textit{beacon request}, it creates a new \textit{beacon report} containing the BSSID, operating frequency, channel number, and RSSI (among other parameters) of each detected AP within its range during the measurement duration specified in the \textit{beacon request}. At the end of the measurement duration, the STA will send a \textit{beacon report} with all the aforementioned gathered information.

Similarly, the \textit{channel load request/report} exchange allows an AP to receive information on the channel condition of a targeted network device. The \textit{channel load report} contains the channel number, actual measurement start time, measurement duration, and channel busy fraction \cite{panaousis2008optimizing}.

%\textcolor{blue}{[Però aquesta STA suposo que necessita un cert temps per prendre les mesures que li demanen, no? Quan envia la resposta?]}
%\textcolor{orange}{Si, aquest temps es l'anomenat 'measurement duration', i s'especifica com a un field mes a la request. La resposta s'envia un cop s'exhaureix el temps corresponent al measurement duration.}

%802.11k: Radio beaconing improvements
%– Neighbor report from AP to client
%– Channel report from AP to client
%– Beacon report from client to AP

\subsection{IEEE 802.11v: Wireless Network Management}

The IEEE 802.11v amendment \cite{5716530} on wireless network management uses network information to influence client roaming decisions. Whereas IEEE 802.11k only targets the radio environment, IEEE 802.11v includes broader operational data regarding network conditions, thus allowing STAs to acquire better knowledge on the topology and state of the network.

In fact, there are a multitude of new services powered by IEEE 802.11v, including power saving mechanisms, interference avoidance mechanisms, fast roaming, or an improved location system, among others. In all cases, the exchange of data among network devices takes place through several action frame formats defined for wireless network management purposes.

%Several Action frame formats are defined for wireless network management (WNM) purposes. A WNM Action field, in the octet field immediately after the Category field, differentiates the formats. The WNM Action field values associated with each frame format are defined in Table 9-354.

The \textit{BSS transition management} service is of special interest to our current work, as it enables to suggest a set of preferred candidate APs to an STA according to a pre-established policy. IEEE 802.11v defines 3 types of \textit{BSS transition management} frames: query, request, and response.
\begin{itemize}
    \item A query is sent by an STA asking for a \textit{BSS transition candidate list} to its corresponding AP.
    \item An AP responds to a query frame with the \textit{BSS transition candidate list}; that is, a request frame containing a prioritized list of preferred APs, their operating frequency, and their channel number, among other information. In fact, the AP may also send a \textit{BSS transition candidate list} to a compatible IEEE 802.11v STA at any time to accelerate any eventual roaming process. 
    \item A response frame is sent by the STA back to the AP, informing whether it accepts or denies the transition.
\end{itemize}

%The 11v BSS Transition Management function allows a wireless distribution system to request a client that supports 11v to roam to an AP that will serve the client better. The client still makes the decision to honor the request or not, but it does offer more control over the roaming process.

Once received a \textit{BSS transition candidate list} and accepted its proposed transition, the STA will follow the provided APs candidate list in order of priority, trying to reassociate to such a network. As operating frequency and channel number of each candidate AP is also provided, total scan process time in the reassociation operation can be minimized. 

%Although the client always has the freedom to choose whether to accept or reject the advice offered by the AP, the additional awareness can assist to build a firm foundation for self-correcting events and actions to be implemented. 

%----------------------------------------------------------------------------------
%----------------------------------------------------------------------------------
%----------------------------------------------------------------------------------
%----------------------------------------------------------------------------------

\section{Channel load aware AP/Extender selection}
\label{load_aware}

We introduce in this section the proposed \textit{channel load aware} AP/Extender selection mechanism. We aim to define a general approach that allows us to study the trade-off between received power and channel load-based metrics to make the AP/Extender selection decision. 

\begin{figure*}[th!]
    \centering
    \includegraphics[width=0.7\textwidth]{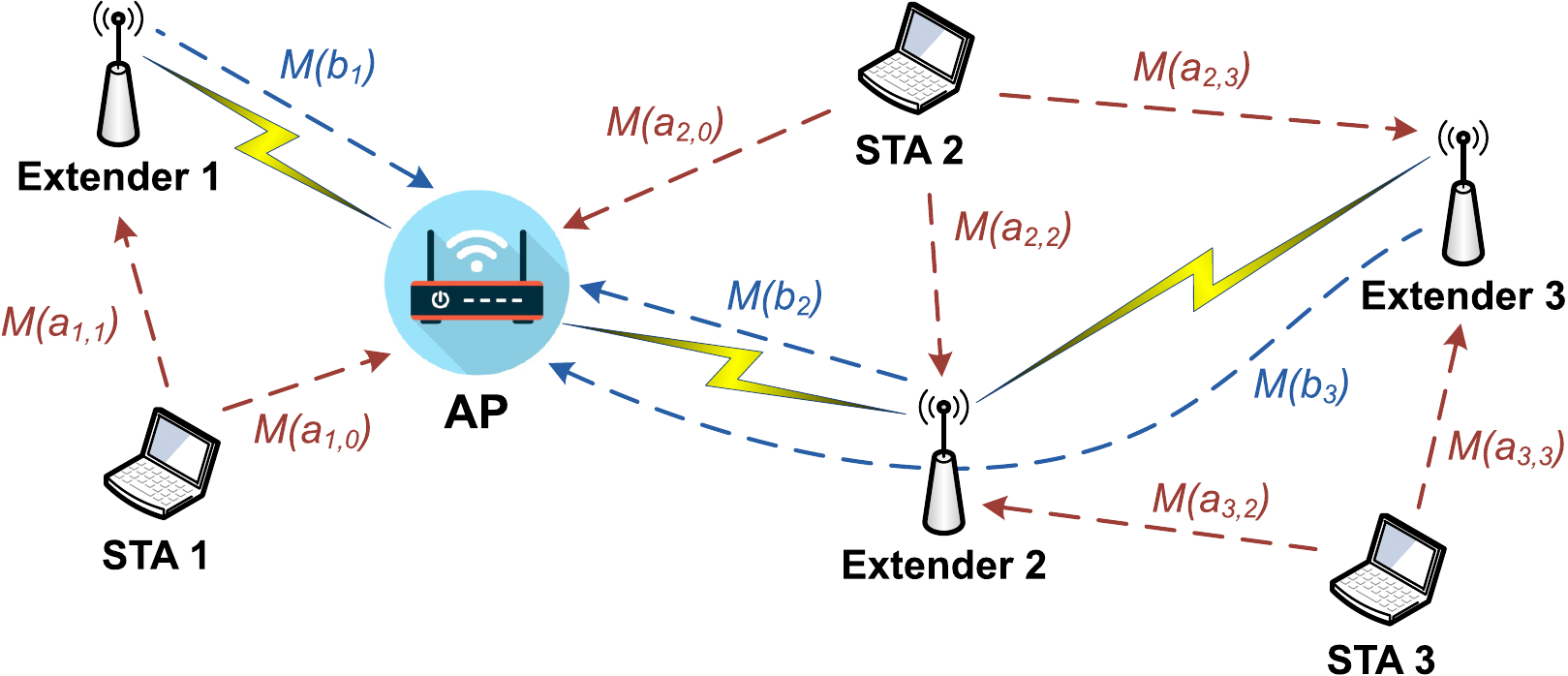}
    \caption{WLAN topology with Extenders. Note that $M(a_{i,j})$ corresponds to the access link metric from STA $i$ to AP/Extender j. As for $M(b_j)$, it corresponds to the backhaul link metric of Extender $j$.}
    \label{fig:topology}
\end{figure*}

The proposed AP/Extender selection mechanism is intended to be applied on a WLAN topology like the one from Figure \ref{fig:topology}, consisting of an AP, several Extenders wirelessly connected to the AP, and multiple STAs willing to associate to the network.\footnote{If Extenders were connected to the AP by means of wired links, the proposed \textit{channel load aware} mechanism would be likewise applicable.}
It is fully based on the existing IEEE 802.11k/v amendments, which enables its real implementation, and can be executed as part of the association process of an STA in any of the following circumstances:

\begin{itemize}
    \item An STA has just associated to the network through the AP/Extender selected by using the default \textit{RSSI-based} criteria.
    \item An STA is performing a roaming procedure between different AP/Extenders from the same WLAN.
    \item The AP initiates an operation to reassociate all previously associated STAs in case network topology has changed (e.g., a new Extender is connected), or an overall load balance operation is executed (e.g., as consequence of new traffic demands coming from STAs).
\end{itemize}

In a real implementation, all computation associated with this mechanism would be executed in the AP, as it is the single, centralized element in the architecture with a global vision of the network. Alternatively, computation tasks could be assumed by an external network controller run into a server, either directly connected to the AP or placed in a remote, cloud-based location.

\subsection{Operation of the AP/Extender selection mechanism}

The \textit{channel load aware} AP/Extender selection mechanism splits the selection process into four differentiated stages. Figure \ref{fig:sequence} shows the sequence of their main tasks, which are described in the following lines:

\begin{figure}[th!]
    \includegraphics[width=0.5\textwidth]{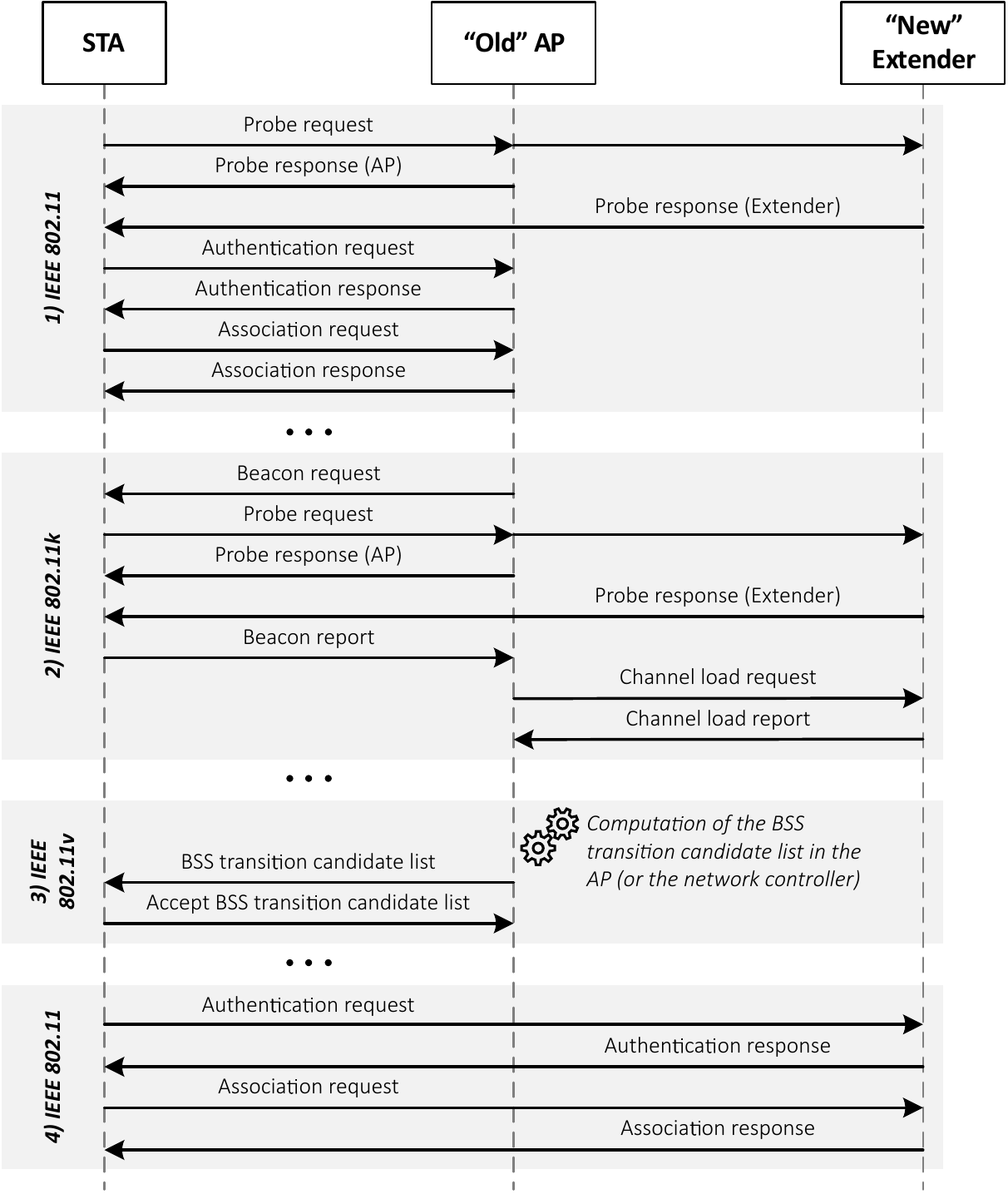}
    \caption{Sequence diagram of the \textit{channel load aware} AP/Extender selection mechanism, using \textit{active scanning} as measurement mode and explicit \textit{channel load request/report} exchange.}
    \label{fig:sequence}
\end{figure}

\begin{enumerate}
    \item Initial association (IEEE 802.11)
    \begin{itemize}
        \item After an \textit{active} or \textit{passive} scanning, the STA sends an association request to the AP/Extender with the best observed RSSI value.
        \item The AP/Extender registers the new STA and confirms its association. Moreover, it checks if the STA supports IEEE 802.11k and IEEE 802.11v modes, which are indispensable to properly perform the next steps of the mechanism.
        \item The AP/Extender notifies the AP (or the network controller) of the new associated STA and its capabilities.
    \end{itemize}
    \item Collection and exchange of information (IEEE 802.11k)
    \begin{itemize}
        \item The AP (or the network controller) initiates a new information collection stage by sending (directly or through the corresponding Extenders) a \textit{beacon request} to the STA.
        \item Depending on the type of the \textit{beacon request} received, the STA initiates an \textit{active scanning}, a \textit{passive scanning}, or simply consults its own \textit{beacon table}.
        \item The surrounding AP/Extenders respond to an \textit{active scanning} with a \textit{probe response} or simply emit their own \textit{beacon frames}.
        \item The STA transmits the resulting \textit{beacon report} to its corresponding AP/Extender.
        \item The AP/Extender, in turn, retransmits this \textit{beacon report} to the AP (or the network controller).
        \item Lastly, the AP emits \textit{channel load requests} to the network Extenders.
        \item The Extenders measure the observed channel occupation and send a \textit{channel load report} to the AP.
    \end{itemize}
    \item Computation and transmission of decision (IEEE 802.11v)
    \begin{itemize}
        \item The AP (or the network controller) computes the $Y_{i,j}$ decision metric (defined in the next subsection) for each AP/Extender detected by the STA.
        \item The AP sends the STA the resulting \textit{BSS transition candidate list}.
    \end{itemize}
    \item Reassociation (IEEE 802.11)
    \begin{itemize}
        \item The STA starts a new association process with the first AP/Extender recommended in the \textit{BSS transition candidate list}. If it fails, the STA tries to associate to the next AP/Extender in the list.
        \item The new AP/Extender registers the new STA and confirms its association.
        \item The new AP/Extender notifies the AP of the new associated STA.
        \item Every reassociation to a new AP/Extender within the WLAN would require a complete authentication process, unless the \textit{fast BSS transition} feature from IEEE 802.11r is employed \cite{bangolae2006performance}. 
    \end{itemize}
\end{enumerate}

%IEEE 802.11r also enables faster roaming by allowing encryption keys to be stored on network APs so the client does not need to perform the complete authentication process every time it roams to a new AP within the network.

\begin{figure}[th!!]
\begin{subfigure}[]{0.48\textwidth}
       \centering \includegraphics[width=\textwidth]{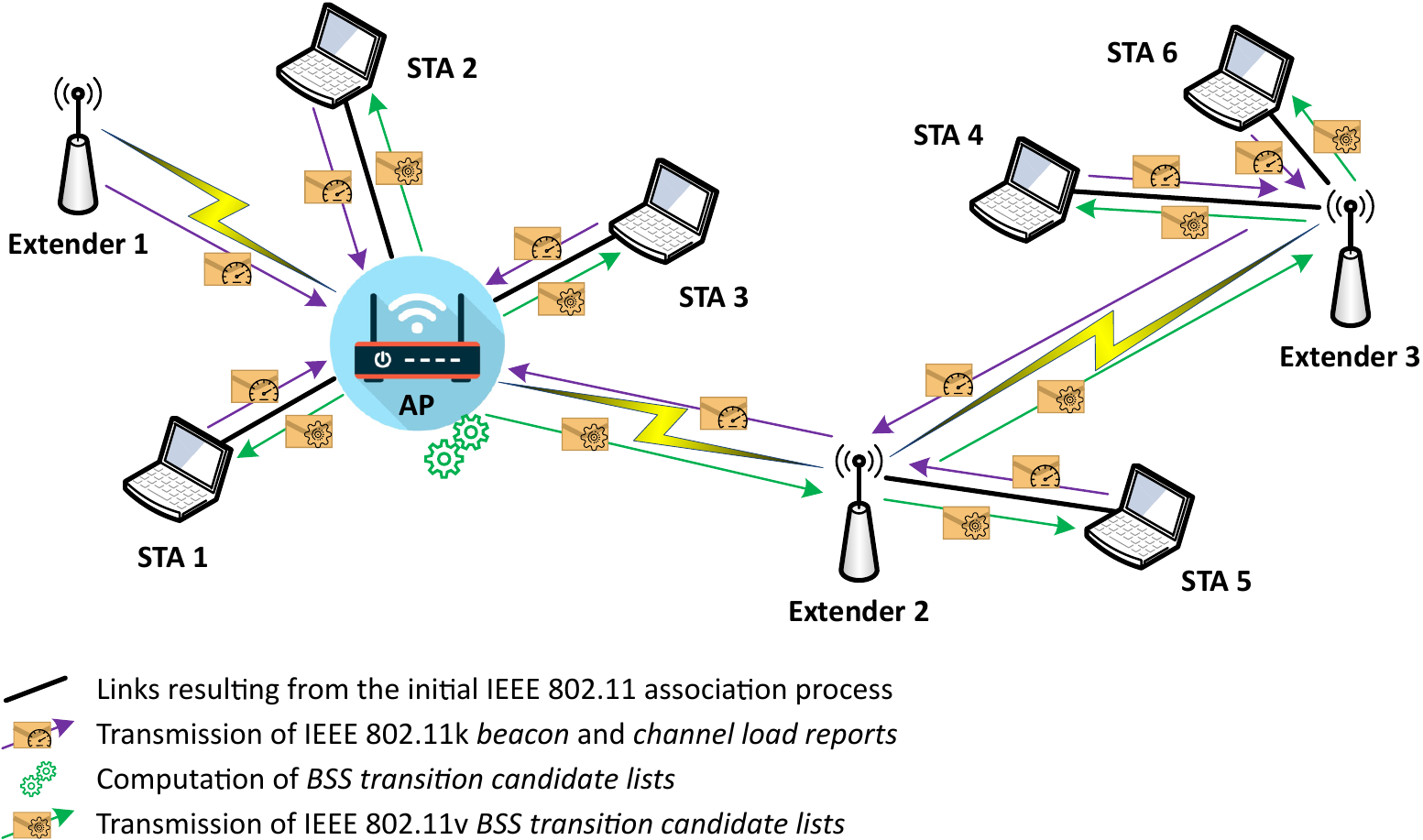}
         \caption{Initial WLAN topology and exchange of network information.}
         \vspace{3mm}
       \label{fig:diag_WLAN_a}
\end{subfigure}
\begin{subfigure}[]{0.48\textwidth}

       \centering \includegraphics[width=\textwidth]{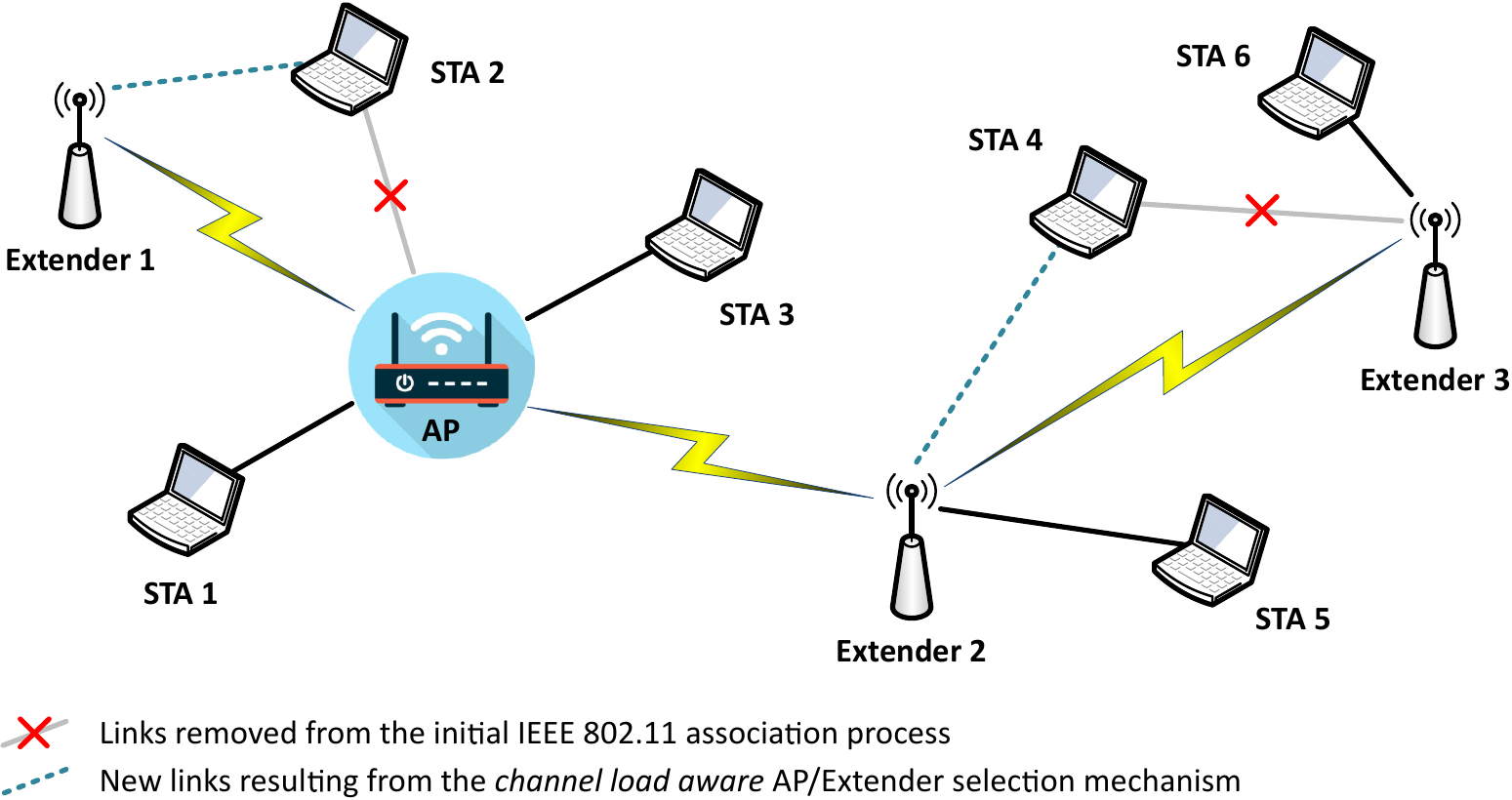}
        \caption{Final WLAN topology.}
        \vspace{3mm}
        \label{fig:diag_WLAN_b}
\end{subfigure} 

\caption{Example of a WLAN before and after applying the \textit{channel load aware} AP/Extender selection mechanism.}
\label{fig:diag_WLAN}
\end{figure}

According to the classification criteria from Table \ref{tab:classification}, the AP selection mode in this new AP/Extender selection mechanism is \textit{hybrid}, because STAs share with the AP information about the network state, the architecture is \textit{centralized}, as the AP (or the network controller) computes the best AP/Extender for each STA, and the parameters of the decision metric are: the RSSI observed by the STA and the channel load observed by the different AP/Extenders. As a matter of example, Figure \ref{fig:diag_WLAN} offers a graphical view of a complete WLAN before and after applying the \textit{channel load aware} AP/Extender selection mechanism.
    
\subsection{AP/Extender selection metric}

The decision metric used in the proposed approach combines parameters observed both in the access link $M(a_{i,j})$ (i.e., from STA $i$ to AP/Extender $j$) and in the backhaul link(s) $M(b_j)$ (i.e., those in the route from Extender $j$ to the AP)~\cite{luo2008improving}.

When using the \textit{RSSI-based} AP selection mechanism, STAs simply choose the AP/Extender with the strongest RSSI value in the access link. Differently, our AP/Extender selection mechanism takes advantage of the capabilities offered by IEEE 802.11k and IEEE 802.11v to create a new decision metric by combining parameters from both access and backhaul links.

More specifically, $Y_{i,j}$ is the decision metric employed in our proposal per each pair formed by STA $i$ and AP/Extender $j$. Then, the best AP/Extender for STA $i$ will be the one with the minimum $Y_{i,j}$ value according to
\begin{eqnarray}
\label{eq:score}
Y_{i,j} &=& \alpha \cdot M(a_{i,j}) + (1-\alpha)\cdot M(b_{j}) =\\ \nonumber
&=& \alpha \cdot \left(\text{RSSI}_{i,j}^*+C_{a_{i,j}} \right)+(1-\alpha)\cdot\sum_{k \in N_{j}} C_{b_{j}}(k), \end{eqnarray}
where $\alpha$ is a configurable factor that weights the influence of access and backhaul links ($0 \leq \alpha \leq 1$) and $C_{a_{i,j}}$ is the channel load of the access link observed by AP/Extender $j$. Considering $N_{j}$ as the set of backhaul links in the path between Extender $j$ and the AP, $C_{b_{j}}(k)$ is the channel load of backhaul link $k$. Note that when $j$ corresponds to the AP, there are no backhaul links (i.e., $N_j = \varnothing$).

Channel load $C$ is here considered as the fraction of time during which the wireless channel is sensed busy, as indicated by either the physical or virtual carrier sense mechanism, with $0 \leq C \leq 1$ \cite{thorpe2008analisys}. The AP (or the network controller) can obtain this information explicitly (by means of the \textit{channel load request/report} exchange) or implicitly (from the \textit{BSS load} element contained in both \textit{beacon frames} and \textit{probe responses} emitted by AP/Extenders) \cite{4544755}.

In fact, unlike other parameters employed in alternative decision metrics, the channel load is able to provide information not only from the targeted WLAN, but also from the influence of other external networks. In consequence, the WLAN is more able to balance the traffic load of newly associated STAs to the less congested AP/Extenders, thus increasing the adaptability degree to the state of the frequency channel.

%BSS Load Element
%The BSS Load element contains information on the current STA population and traffic levels in the BSS.
%The element information format is defined in Figure 9-260. This element might be used by the STA for vendor-specific AP selection algorithm when roaming.

%The Channel Utilization field is defined as the percentage of time, normalized to 255 linearly scaled with 255 representing 100\%, that the AP sensed the medium was busy, as indicated by either the physical or virtual carrier sense (CS) mechanism.

For its part, $\text{RSSI}_{i,j}^*$ corresponds to an inverse weighting of the signal strength received by STA $i$ from AP/Extender $j$, which is computed as
\begin{align}
\text{RSSI}_{i,j}^{*} = \frac{\text{RSSI}_{i,j}-P_{t_{j}}}{S_i-P_{t_{j}}},
\label{eq:rssi}
\end{align}
where $\text{RSSI}_{i,j}$ is the signal strength received by STA $i$ from AP/Extender $j$ in dBm, $P_{t_{j}}$ is the transmission power level of AP/Extender $j$ in dBm, and $S_i$ is the carrier sense threshold (i.e., sensitivity level) of STA $i$ in dBm.

 \begin{figure}[h]
    \centering
    \includegraphics[width=0.40\textwidth]{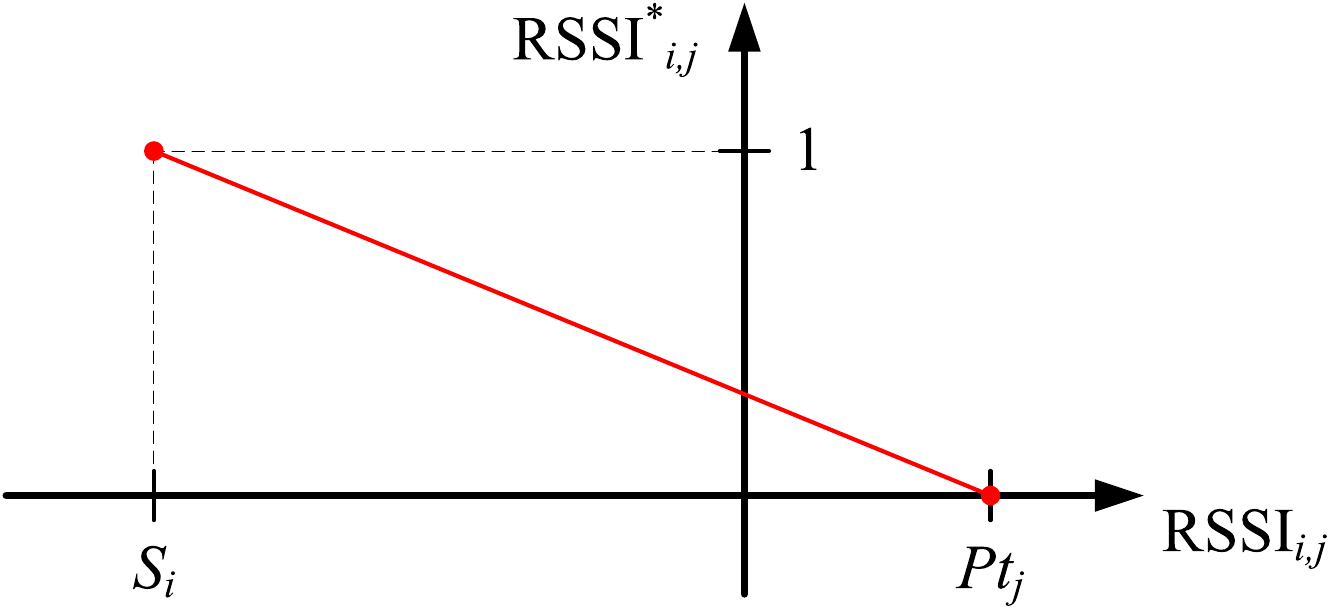}
    \caption{RSSI weighting applied in the \textit{channel load aware} AP/Extender selection mechanism.}
    \label{fig:weighting}
\end{figure}

As shown in Figure \ref{fig:weighting}, the weighting of possible input values of $\text{RSSI}_{i,j} \in [S_{i},P_{t_{j}}]$ from (\ref{eq:rssi}) applied in the AP/Extender selection mechanism results in output values of $\text{RSSI}_{i,j}^{*} \in [0,1]$. Consequently, low RSSI values (i.e., those close to the sensitivity level $S_{i}$) are highly penalized.

%----------------------------------------------------------------------------------
%----------------------------------------------------------------------------------
%----------------------------------------------------------------------------------
%----------------------------------------------------------------------------------

\section{Performance evaluation}
\label{simulation}

This section is first intended to understand the benefits of adding Extenders to a WLAN, and determine their optimal number and location for a given area. Then, the very concept of a WLAN with Extenders is applied to a typical Home WiFi scenario aiming to evaluate the impact of the main parameters involved in the AP/Extender selection mechanism on network's performance.

\subsection{Simulation framework}

MATLAB was the selected tool to develop a simulator that enables the deployment, setting, testing, and performance evaluation of a WLAN. Specifically, our simulator focused on the AP/Extender selection mechanism contained in the STA association process, the transmission of uplink (UL) data packets (i.e., those from STAs to the AP), and the computation of metrics in the AP with respect to the received traffic.

As for the PHY layer, it was assumed that, once the network topology was established, all devices adjusted their data rate according to the link condition. Specifically, simulations used the ITU-R indoor site-general path loss model according to \begin{eqnarray}
\text{PL}_{\text{ITU}}(d_{i,j}) = 20 \cdot \log_{10}(f_{c}) + N \cdot \log_{10}(d_{i,j}) + L_{f} - 28,
\label{eq:model_itu}
\end{eqnarray}where $\text{PL}_{\text{ITU}}$ is the path loss value (in dB), $d_{i,j}$ is the distance between transmitter $i$ and receiver $j$ (in m), $f_{c}$ is the employed frequency (in MHz), $N$ is the distance power loss coefficient (in our particular case and according to the model guidelines, $N = 31$), and $L_{f}$ is the floor penetration loss factor (which was removed as a single floor was always considered)  \cite{itu2012p}.

The distributed coordination function (DCF) was used by all AP/Extenders and STAs. We assumed that all AP/Extenders and STAs were within the coverage area of the others, given they operated in the same channel. Therefore, an STA was able to associate to any AP/Extender in the area of interest. 

Only UL transmissions were considered in simulations, as they represent the worst case in a WLAN; that is, when multiple non-coordinated devices compete for the same wireless spectrum. Though excluded from the current study, downlink (DL) communications could either follow the same topology resulting from the STA association process or, as it is already conceived by designers of future WiFi 7, establish their own paths by means of the multi-link operation capability (in our particular case, according to an alternative decision metric) \cite{adame2019time}.

WLAN performance metrics (throughput, delay, and congestion) were obtained using the IEEE 802.11 DCF model presented and validated in \cite{bellalta2005simple}, which supports heterogeneous finite-load traffic flows as required in this work. Details from two different wireless standards were implemented in the simulator: IEEE 802.11n and IEEE 802.11ac. Due to the higher penetration of 2.4 GHz compatible devices in real deployments, all tests employed IEEE 802.11n at 2.4 GHz in access links (with up to 3 available orthogonal channels) and IEEE 802.11ac at 5 GHz in backhaul links (with a single channel).\footnote{Data rates were computed from the observed RSSI and according to the corresponding modulation and coding scheme (MCS) table.} Nonetheless, the simulator supports any combination of standards over the aforementioned network links.

%L'Alvaro ha fet unes simulacions amb ns3 que permeten validar el model. Per si fan falta, es podrien posar com appendix.

\begin{table}[t!!]
\centering
\footnotesize
\caption{List of common simulation parameters.}
\label{tab:parameters}
\begin{tabular}{|c|c|c|c|c|}
\hline
                                                                           \textbf{Parameter}                                              & \textbf{Symbol}     & \textbf{Description}                                                                     & \textbf{Value} & \textbf{Unit} \\ \hline
\multirow{3}{*}{\begin{tabular}[c]{@{}c@{}}\textbf{Operating}\\\textbf{frequency}\end{tabular}}                                 & $f_{a}$               & \begin{tabular}[c]{@{}c@{}}Frequency band \\of access links \end{tabular}                                                             & 2.4            & GHz           \\ \cline{2-5} 
                                                                                                                         & $f_{b}$               & \begin{tabular}[c]{@{}c@{}}Frequency band \\of backhaul links \end{tabular}                                                         & 5              & GHz           \\ \hline
\multirow{3}{*}{\begin{tabular}[c]{@{}c@{}}\textbf{Operating}\\\textbf{channel}\end{tabular}}                                    & $c_{a}$              &
\begin{tabular}[c]{@{}c@{}}Available channels\\in the access link  \end{tabular} 

                                                                 & $\left\lbrace1,6,11\right\rbrace$     & -       \\ \cline{2-5} 
                                                                                                                         & $c_{b}$              & \begin{tabular}[c]{@{}c@{}}Available channels\\in the backhaul link  \end{tabular}                                                               &        $\left\lbrace 36 \right\rbrace$       & -                   \\ \hline
                                                                                                                         
                 \multirow{6}{*}{\textbf{Extenders}}                                                                             & $N_\text{E}$                      & Number of Extenders                                                                            & \textit{variable}         & Extenders          \\ \cline{2-5} 
                                                                                                                         & $N_{\text{CE}}$ &                                \begin{tabular}[c]{@{}c@{}}Maximum number\\ of consecutive  \\Extenders in\\ the same path \end{tabular}                          & 2       & Extenders  
                                                                                                                         \\ \cline{2-5} 
                                                                                                                         & $N_{\text{STA}}$ & Number of STAs                                                         & 10       & STAs \\ \hline

\multirow{7}{*}{\begin{tabular}[c]{@{}c@{}}\textbf{Traffic}\\\textbf{generation}\end{tabular}}                                                                             & $L$                      & Packet length                                                                            & 12000         & bits          \\ \cline{2-5} 
                                                                                                                         & $B_{\text{STA}}$ &       
                                                                                                                         \begin{tabular}[c]{@{}c@{}}Traffic load\\ (per STA)  \end{tabular} & \textit{variable}       & bps   \\
                                                                                                                         \cline{2-5} 
                                                                                                                         & $B_T$ &       
                                                                                                                         \begin{tabular}[c]{@{}c@{}}Traffic load\\(total network)  \end{tabular} & \textit{variable}       & bps    
                                                                                                                         
                     \\
                                                                                                                         \cline{2-5} 
                                                                                                                         & $B_{\text{EXT}}$ &       
                                                                                                                         \begin{tabular}[c]{@{}c@{}}Traffic load\\(external network)  \end{tabular} & \textit{variable}       & bps                                                                                                      \\ \hline \multirow{5}{*}{\begin{tabular}[c]{@{}c@{}}\textbf{Radio}\\\textbf{module}\end{tabular}}                                                                                   & $P_{t}$              &                        \begin{tabular}[c]{@{}c@{}}Transmission\\power level  \end{tabular}                                                           & 20             & dBm           \\ \cline{2-5} 
                                                                                                                         & $S$                   &      
                                                                                                                         \begin{tabular}[c]{@{}c@{}}Receiver's\\sensitivity level  \end{tabular} & -90            & dBm           \\ \cline{2-5} 
                                                                                                                         & $\text{SS}$                     &       
                                                                               \begin{tabular}[c]{@{}c@{}}Number of\\spatial streams  \end{tabular}                                          
                                                                                                                         & 2              & -             \\  \hline
\multirow{4}{*}{\begin{tabular}[c]{@{}c@{}}\textbf{Channel load}\\\textbf{aware AP/Ext}\\ \textbf{selection} \\\textbf{mechanism}\end{tabular}} &      $\alpha$                  & \begin{tabular}[c]{@{}c@{}}Weighting\\ factor\end{tabular} & \textit{variable}       & -            \\  \cline{2-5} 
                                                                                                                                                                & $\beta$  & \begin{tabular}[c]{@{}c@{}}Share of\\ IEEE 802.11k/v\\  capable STAs\end{tabular}                                                                         & \textit{variable}       & \%                                                                                 
                                                                                                                                        \\  \hline
     
                    \textbf{Deployments}  & $k$ &    \begin{tabular}[c]{@{}c@{}}Random STA\\deployments  \end{tabular}
                    
                    & \textit{variable} & -    \\                                                                            
                                                                                                                 \hline
\end{tabular}
\end{table}

\begin{table*}[]
\centering
\footnotesize
\caption{List of \textit{test-specific} simulation parameters.}
\label{tab:parameters2}
\begin{tabular}{|c|c|c|c|c|c|c|c|c|c|c|c|c|c|c|}
\hline
\multirow{3}{*}{\textbf{Test}}     & \multirow{3}{*}{\begin{tabular}[c]{@{}c@{}}\textbf{Deployment} \\ \textbf{area}\end{tabular}}              & \multirow{3}{*}{$k$} & \multirow{3}{*}{\begin{tabular}[c]{@{}c@{}}$B_{\text{STA}}$ \\ \textbf{(Mbps)}\end{tabular}}                                              & \multirow{3}{*}{$N_{\text{STA}}$} & \multirow{3}{*}{$N_{\text{E}}$} & \multicolumn{5}{c|}{\begin{tabular}[c]{@{}c@{}}\textbf{Channel selected in the}\\ \textbf{access link ($f_a$ = 2.4 GHz)}\end{tabular}}                                    & \multirow{3}{*}{\begin{tabular}[c]{@{}c@{}}\textbf{AP/Extender}\\ \textbf{selection}\\ \textbf{mechanism}\end{tabular}} & \multirow{3}{*}{$\alpha$} & \multirow{3}{*}{$\beta$} & \multirow{3}{*}{\begin{tabular}[c]{@{}c@{}}$B_{\text{EXT}}$ \\ \textbf{(Mbps)}\end{tabular}} \\ \cline{7-11}
                                   &                                                                                                   &                             &                                                                               &                                  &                                & \multirow{2}{*}{\textbf{AP}} & \multirow{2}{*}{\textbf{E\textsubscript{1}}} & \multirow{2}{*}{\textbf{E\textsubscript{2}}} & \multirow{2}{*}{\textbf{E\textsubscript{3}}} & \multirow{2}{*}{\textbf{E\textsubscript{4}}} &                                                                                                       &                                 &                                &                                 \\
                                   &                                                                                                   &                             &                                                                               &                                  &                                &                              &                              &                              &                              &                              &                                                                                                       &                                 &                                &                                 \\ \hline
\multirow{4}{*}{\textbf{Test 1.1}} & \multirow{4}{*}{Circular area}                                                                      & \multirow{4}{*}{1000}      & \multirow{4}{*}{2.4}                                                          & \multirow{4}{*}{10}              & 0, 4                           & 1                            & 6                            & 6                            & 11                           & 11                           & \textit{RSSI-based}                                                                                   & -                               & -                              & -                               \\ \cline{6-15} 
                                   &                                                                                                   &                             &                                                                               &                                  & 4                              & 1                            & 6                            & 6                            & 11                           & 11                           & \textit{Load aware}                                                                                   & 0.5                             & 100                            & -                               \\ \cline{6-15} 
                                   &                                                                                                   &                             &                                                                               &                                  & \multirow{2}{*}{4}             & 1                            & 1                            & 1                            & 1                            & 1                            & \textit{RSSI-based}                                                                                   & -                               & -                              & -                               \\ \cline{7-15} 
                                   &                                                                                                   &                             &                                                                               &                                  &                                & 1                            & 1                            & 1                            & 1                            & 1                            & \textit{Load aware}                                                                                   & 0.5                             & 100                            & -                               \\  \hline
\multirow{2}{*}{\textbf{Test 1.2}} & \multirow{2}{*}{\begin{tabular}[c]{@{}c@{}}Circular area of \\ radius $1.2 \cdot D_{max}$\end{tabular}}                                                                      & \multirow{2}{*}{10000}      & \multirow{2}{*}{2.4}                                                  & \multirow{2}{*}{10}              & 0, 2, 4                        & 1                            & 6                            & 6                            & 11                           & 11                           & \textit{RSSI-based}                                                                                   & -                               & -                              & -                               \\ \cline{6-15} 
                                   &                                                                                                   &                             &                                                                               &                                  & 2, 4                           & 1                            & 6                            & 6                            & 11                           & 11                           & \textit{Load aware}                                                                                   & 0.5                             & 100                            & -                               \\ \hline
\multirow{2}{*}{\textbf{Test 1.3}} & \multirow{2}{*}{Circular area}                                                                      & \multirow{2}{*}{1000}      & \multirow{2}{*}{{[}0.012, 3.6{]}}                                                  & \multirow{2}{*}{10}              & 0, 2, 4                        & 1                            & 6                            & 6                            & 11                           & 11                           & \textit{RSSI-based}                                                                                   & -                               & -                              & -                               \\ \cline{6-15} 
                                   &                                                                                                   &                             &                                                                               &                                  & 2, 4                           & 1                            & 6                            & 6                            & 11                           & 11                           & \textit{Load aware}                                                                                   & 0.5                             & 100                            & -                               \\\hline
\multirow{4}{*}{\textbf{Test 2.1}} & \multirow{4}{*}{Home WiFi}          & \multirow{4}{*}{1000}      & \multirow{4}{*}{{[}0.012, 6{]}}                                                  & \multirow{4}{*}{10}              & 0, 1, 2                        & 1                            & 6                            & 11                           & -                            & -                            & \textit{RSSI-based}                                                                                   & -                               & -                              & -                               \\ \cline{6-15} 
                                   &                                                                                                   &                             &                                                                               &                                  & 1, 2                           & 1                            & 6                            & 11                           & -                            & -                            & \textit{Load aware}                                                                                   & 0.5                             & 100                            & -                               \\ \cline{6-15} 
                                   &                                                                                                   &                             &                                                                               &                                  & 1, 2                           & 1                            & 1                            & 1                            & -                            & -                            & \textit{RSSI-based}                                                                                   & -                               & -                              & -                               \\ \cline{6-15} 
                                   &                                                                                                   &                             &                                                                               &                                  & 1, 2                           & 1                            & 1                            & 1                            & -                            & -                            & \textit{Load aware}                                                                                   & 0.5                             & 100                            & -                               \\ \hline
\multirow{2}{*}{\textbf{Test 2.2}} & \multirow{2}{*}{Home WiFi}         & \multirow{2}{*}{1000}      & \multirow{2}{*}{\begin{tabular}[c]{@{}c@{}}1.8, 3,\\ 4.2, 5.4\end{tabular}} & \multirow{2}{*}{10}              & \multirow{2}{*}{2}             & 1                            & 6                            & 11                           & -                            & -                            & \textit{Load aware }                                                                                  & {[}0,1{]}                       & 100                            & -                               \\ \cline{7-15} 
                                   &                                                                                                   &                             &                                                                               &                                  &                                & 1                            & 1                            & 1                            & -                            & -                            &                                  \textit{Load aware }                                                    & {[}0,1{]}                       & 100                            & -                               \\ \hline
\multirow{2}{*}{\textbf{Test 2.3}} & \multirow{2}{*}{Home WiFi}          & \multirow{2}{*}{1000}      & \multirow{2}{*}{\begin{tabular}[c]{@{}c@{}}1.8, 3,\\ 4.2, 5.4\end{tabular}} & \multirow{2}{*}{10}              & \multirow{2}{*}{2}             & 1                            & 6                            & 11                           & -                            & -                            & \textit{Load aware}                                                                                   & 0.5                             & {[}0,100{]}                    & -                               \\ \cline{7-15} 
                                   &                                                                                                   &                             &                                                                               &                                  &                                & 1                            & 1                            & 1                            & -                            & -                            & \textit{Load aware}                                                                                   & 0.5                             & {[}0,100{]}                    & -                               \\\hline
                                   \multirow{2}{*}{\textbf{Test 2.4}} & \multirow{2}{*}{Home WiFi}       & \multirow{2}{*}{1}      & \multirow{2}{*}{4.32} & \multirow{2}{*}{10}              & 0,1             & 1                            & 6                            & -                           & -                            & -                            & \textit{RSSI-based}                                                                                   & -                               & -                    & {[}0,12{]}                               \\ \cline{6-15} 
                                   &                                                                                                   &                             &                                                                               &                                  &                1                & 1                            & 6                            & -                            & -                            & -                            & \textit{Load aware}                                                                                   & {[}0.5,1{]}                               & 100                   & {[}0,12{]}                               \\\hline
\end{tabular}
\end{table*}

A wide set of tests was conducted on several predefined scenarios to evaluate the impact of different WLAN topologies, configurations, and AP/Extender selection mechanisms on the network's performance. The definition of the scenarios together with their corresponding tests is provided in the following subsections. Lastly, a comprehensive list of common simulation parameters is offered in Table \ref{tab:parameters}, whose values were applied to all subsequent tests, if not otherwise specified. As for \textit{test-specific} simulation parameters, we refer the reader to Table \ref{tab:parameters2}. 

\subsection{Scenario \#1: Circular area}

\begin{figure}[h!!!]
    \centering
    \begin{subfigure}[t]{0.2\textwidth}
          \centering \includegraphics[width=\textwidth]{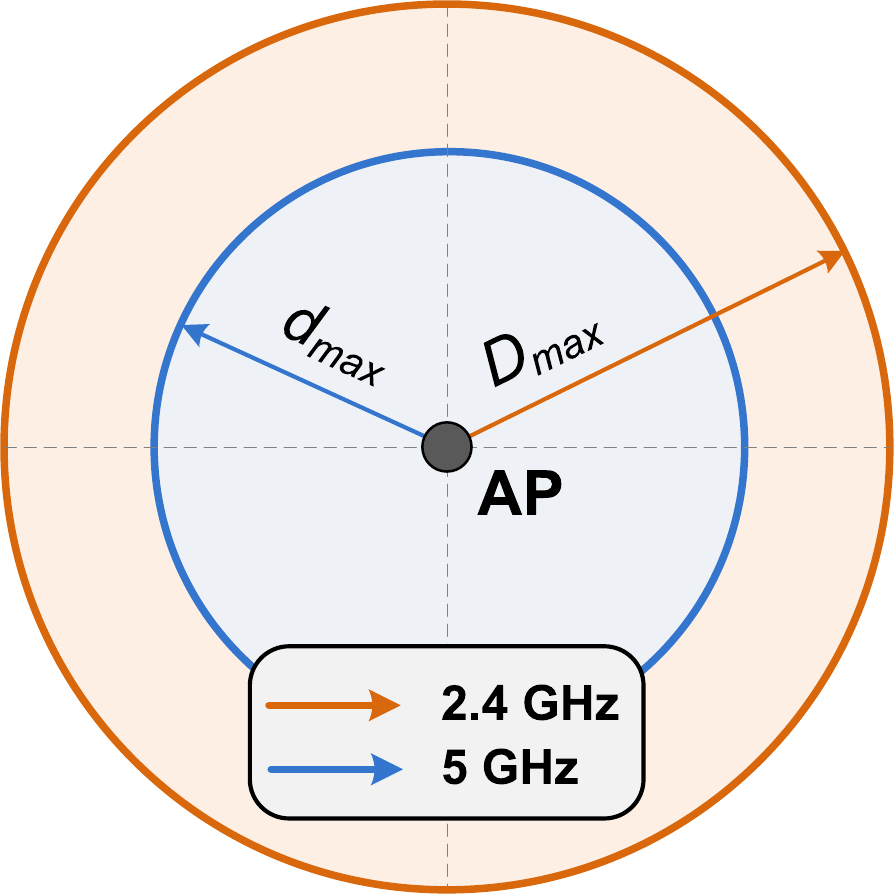}
             \caption{Without Extenders.}
                  \vspace{3mm} \label{fig:scenario1a}
    \end{subfigure}
  
    \begin{subfigure}[t]{0.2\textwidth}
        \centering \includegraphics[width=\textwidth]{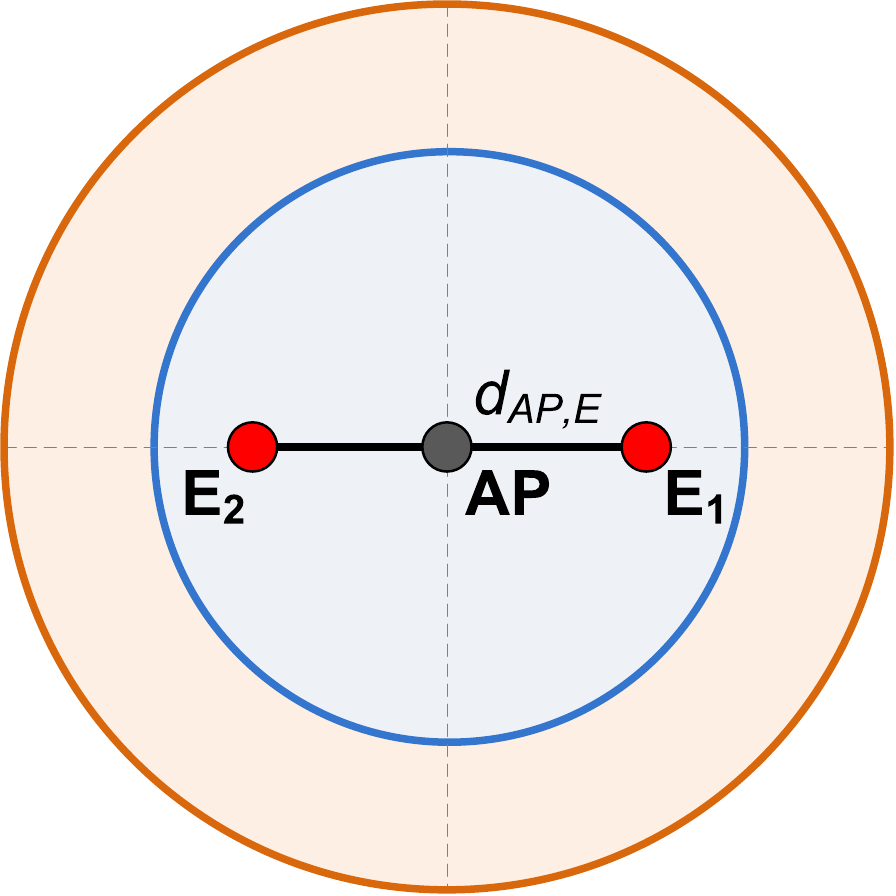}
        \caption{With 2 Extenders.}
        \label{fig:scenario1c}
    \end{subfigure} 
    \begin{subfigure}[t]{0.2\textwidth}
        \centering \includegraphics[width=\textwidth]{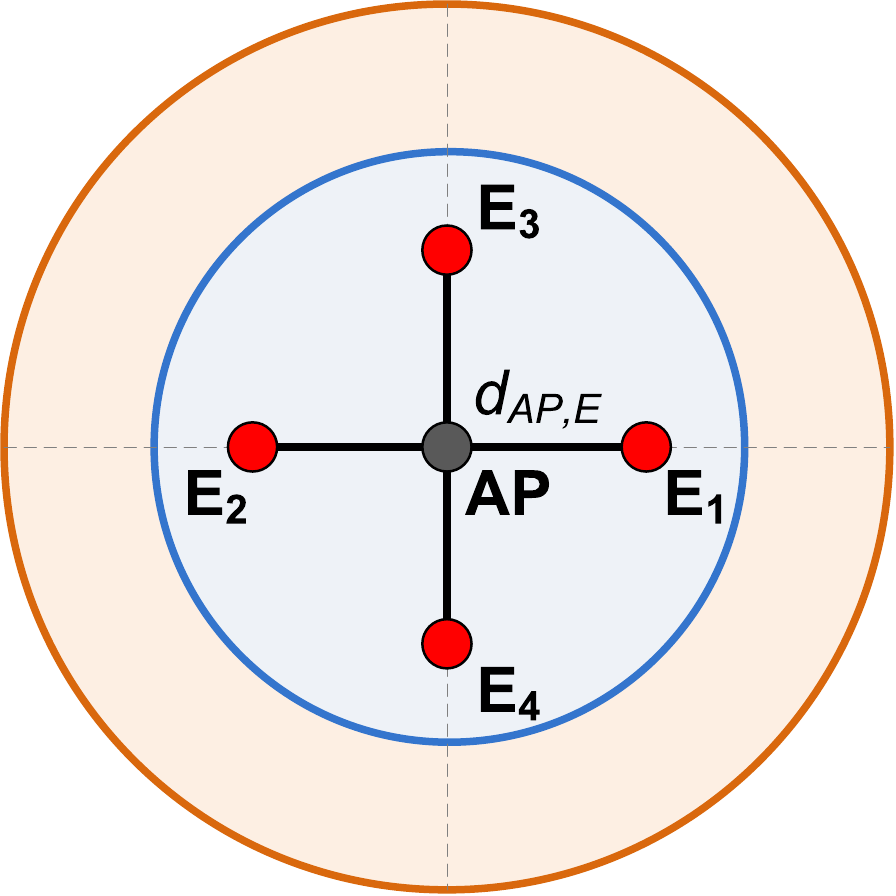}
        \caption{With 4 Extenders.}
        \label{fig:scenario1d}
    \end{subfigure} 
    \caption{Network topologies of Scenario \#1.}
    \label{fig:scenario1}
\end{figure}

A circular area was defined by the maximum coverage range of the AP at 2.4 GHz ($D_{max}$); i.e., the distance in which an STA would receive a signal with the same strength as its sensitivity level. Three different network topologies were there considered: only a single AP, an AP and 2 Extenders, and an AP and 4 Extenders forming a cross (see Figure \ref{fig:scenario1}). Position of Extenders was in turn limited by the maximum coverage range of the AP at 5 GHz ($d_{max}$). 

\subsubsection{Test 1.1: AP-Extender distance}

The goal of this test was to evaluate the effect of the distance between the AP and any Extender ($d_{\text{AP,E}}$) on network's performance. To keep symmetry, the topology from Figure~\ref{fig:scenario1d} was used, moving all Extenders far from the AP, with RSSI values at any Extender ($\text{RSSI}_{\text{AP,E}}$) ranging from $-50$ dBm to $-90$ dBm (i.e., being the latter the $\text{RSSI}_{\text{AP,E}}$ value at $d_{max}$), in intervals of 1 dB. The case without Extenders was also included for comparative purposes.

A number of $N_{\text{STA}} = $ 10 STAs with a common traffic load of $B_{\text{STA}} = $ 2.4 Mbps were uniformly and randomly deployed $k =$ 1000 times on the AP coverage area. Both the \textit{RSSI-based} and the \textit{channel load aware} AP/Extender selection mechanisms were used in each deployment. In the latter case, $\alpha$ was set to 0.5 to give the same importance to access and backhaul links when selecting an AP/Extender.

\begin{figure*}[t!!!]
    \centering
    \includegraphics[width=0.75\textwidth]{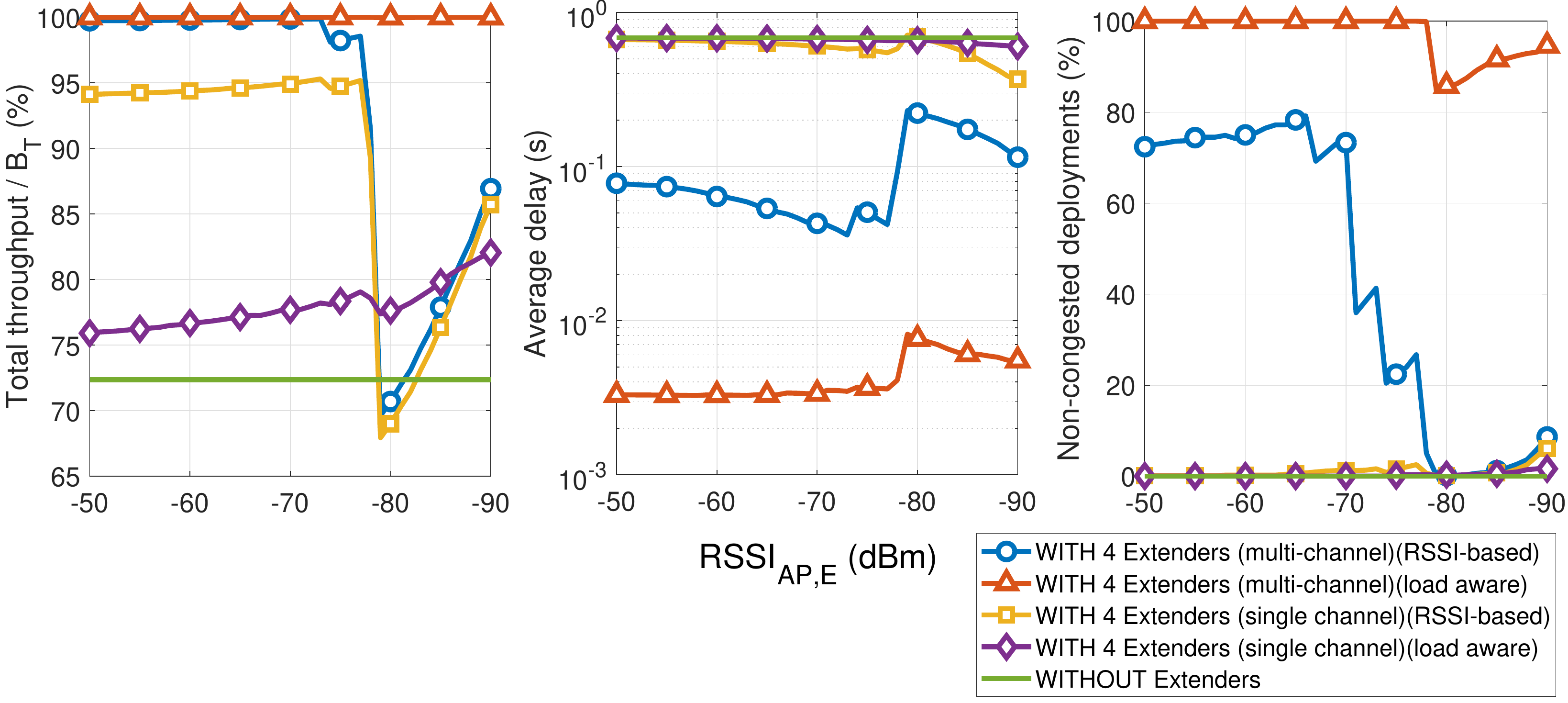}
    \caption{Test 1.1. AP-Extender distance.}
    \label{fig:1_1_mix}
\end{figure*}

As shown in Figure \ref{fig:1_1_mix}, the use of Extenders almost always improved the network's performance in terms of throughput, delay, and congestion regardless $\text{RSSI}_{\text{AP,E}}$. In general, the best range to place Extenders was $\text{RSSI}_{\text{AP,E}} \in \left [-50, -72 \right ]$ dBm, as throughput was maintained over 99\% in multi-channel cases when using any of the analyzed AP/Extender selection mechanisms.\footnote{In this test, but also as generalized practice in the rest of tests from this article, results of each network configuration were obtained as the mean of values from all $k$ deployments, whether the network got congested or not.}

\begin{figure*}[t!!]
    \centering
    \includegraphics[width=0.75\textwidth]{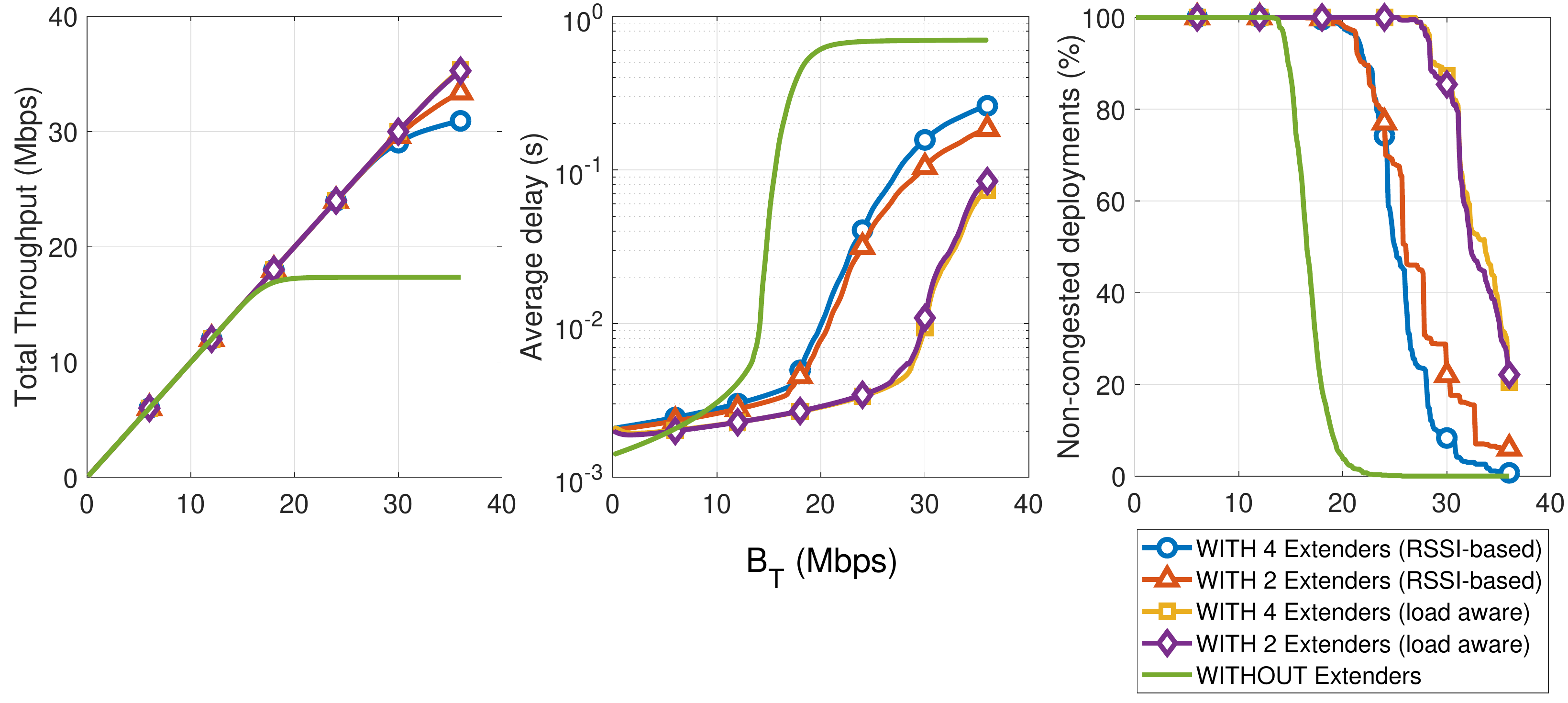}
    \caption{Test 1.3. Number of Extenders.}
    \label{fig:1_3_mix}
\end{figure*}

More specifically, the \textit{channel load aware} mechanism was able to ensure 100\% of throughput and keep delay below 10~ms regardless $\text{RSSI}_{\text{AP,E}}$. This was not the case when using a single communication channel, because almost all STAs were directly connected to the AP (thus resembling the case without Extenders, where furthest STAs hindered the operation of the rest due to their higher channel occupancy), unless they were really close to an alternative Extender. 

As for the \textit{RSSI-based} mechanism, it always behaved worse than the \textit{channel load aware} mechanism in multi-channel cases, but provided better performance in single channel ones. In fact, although the number of STAs connected to Extenders decayed as we moved Extenders far away from the AP, that value was still much higher than in the \textit{channel load aware} mechanism. However, the adoption by Extenders of MCS 1 from $\text{RSSI}_{\text{AP,E}}$ = -77 dBm on, severely impacted on network's performance, as they were not able to appropriately transmit all packets gathered from STAs.

As a result of this test, $d_{\text{AP,E}}$ was set in following tests to the value that made $\text{RSSI}_{\text{AP,E}}$ = -70 dBm.

\subsubsection{Test 1.2: Network's range extension}

To prove the benefit of using Extenders to increase the network coverage, the same topologies of Scenario~\#1 were used. However, in this case, STAs were placed uniformly at random over a circular area of radius $1.2 \cdot D_{max}$. Again,  \textit{RSSI-based} and \textit{channel load aware} (with $\alpha =$ 0.5) AP/Extender selection mechanisms were employed.

A number of $N_{\text{STA}} = $ 10 STAs were randomly deployed $k=$ 10000 times on the predefined area, with the resulting average rate of successful associations from Table \ref{tab:association_1_1}. As expected, the higher the number of Extenders, the higher the total percentage of STAs that found an AP/Extender within their coverage area and got associated. In fact, both AP/Extender selection mechanisms achieved the same STA association rates, because they only depended on whether there were available AP/Extenders within each STA coverage area.

\begin{table}[h]
\centering
\footnotesize
\caption{Test 1.2. Network's range extension.}
\label{tab:association_1_1}
\begin{tabular}{|c|c|}
\hline
\textbf{Network configuration}      & \textbf{Associated STAs} \\ \hline
WITH 4 Extenders (\textit{RSSI-based}) & 93.432\%                  \\ \hline
WITH 4 Extenders (\textit{load aware}) & 93.432\%                  \\ \hline
WITH 2 Extenders (\textit{RSSI-based}) & 90.330\%                  \\ \hline
WITH 2 Extenders (\textit{load aware}) & 90.330\%                  \\  \hline
WITHOUT Extenders           & 83.489\%                  \\ \hline
\end{tabular}
\end{table}

\begin{table*}[h!!!!]
\centering
\footnotesize
\caption{Test 1.3. Number of Extenders (network's operational range expressed in terms of $B_T$).}
\label{tab:results_1_3}
\begin{tabular}{|l|c|c|c|}
\hline
\multirow{2}{*}{\textbf{Network configuration}} & \multicolumn{3}{c|}{$B_T$ \textbf{(Mbps)}}                                                                                                                                                                                  \\ \cline{2-4} 
                                                & \textbf{Throughput $\geqslant$ 99\%} & 
                                                
          \textbf{Delay} $\leqslant$ \textbf{10 ms}                        & \begin{tabular}[c]{@{}c@{}}\textbf{No congested}\\ \textbf{deployments} \end{tabular} \\ \hline
\textbf{WITH 4 Extenders (RSSI-based)}          & [0, 28.20]                                                                                                     & [0, 20.04]                            & [0, 17.16]                                                                                       \\ \hline
\textbf{WITH 2 Extenders (RSSI-based)}          & [0, 29.40]                                                                                                     & [0, 20.76]                             & [0, 16.44]                                                                                       \\ \hline
\textbf{WITH 4 Extenders (load aware)}          & [0, 34.56]                                                                                                     & [0, 30.12]                             & [0, 27.12]                                                                                       \\ \hline
\textbf{WITH 2 Extenders (load aware)}          & [0, 34.32]                                                                                                     & [0, 29.88]                             & [0, 25.44]                                                                                       \\ \hline
\textbf{WITHOUT Extenders}                      & [0, 15.96]                                                                                                     & [0, 14.16]                             & [0, 13.20]                                                                                       \\ \hline
\end{tabular}
\end{table*}

\subsubsection{Test 1.3: Number of Extenders}

In all three topologies from Scenario \#1 were placed a number of $N_{\text{STA}} =$ 10 STAs, each one with the same traffic load ranging from $B_{\text{STA}} = 12$ kbps to $B_{\text{STA}} = 3.6$ Mbps (i.e., a total network traffic, $B_T = N_{\text{STA}} \cdot B_{\text{STA}}$, from $B_T = 0.12$ Mbps to $B_T = 36$ Mbps). STA deployments were randomly selected $k =$ 1000 times and the whole network operated under both the \textit{RSSI-based} and the \textit{channel load aware} (with $\alpha =$ 0.5) AP/Extender selection mechanisms. In this test, only the multi-channel case was considered.

Results from Figure \ref{fig:1_3_mix} justify the use of Extenders to increase the range in which the network operates without congestion, going up to $B_T \approx 13$ Mbps without Extenders, up to $B_T \approx 16$ Mbps in the \textit{RSSI-based} mechanism, and up to $B_T \approx 25$ Mbps in the \textit{channel load aware} one. Furthermore, the \textit{channel load aware} mechanism guaranteed the minimum observed delay for any considered value of $B_T > 5$ Mbps.

The influence of the number of Extenders on performance was different in function of the AP/Extender selection mechanism. Whereas it was barely relevant in the \textit{channel load aware} mechanism due to the effective load balancing among Extenders and AP, it provided heterogeneous results when using the \textit{RSSI-based} mechanism. Particularly, the use of 4 Extenders left the AP with a very low number of directly connected STAs, thus overloading backhaul links with respect to the case with only 2 Extenders. Lastly, further details on network's operational range are detailed in Table \ref{tab:results_1_3} according to three different metrics based on throughput, delay, and congestion.
    
\subsection{Scenario \#2: Home WiFi}

\begin{figure}[th!!]
\begin{subfigure}[]{0.45\textwidth}
       \centering \includegraphics[width=0.75\textwidth]{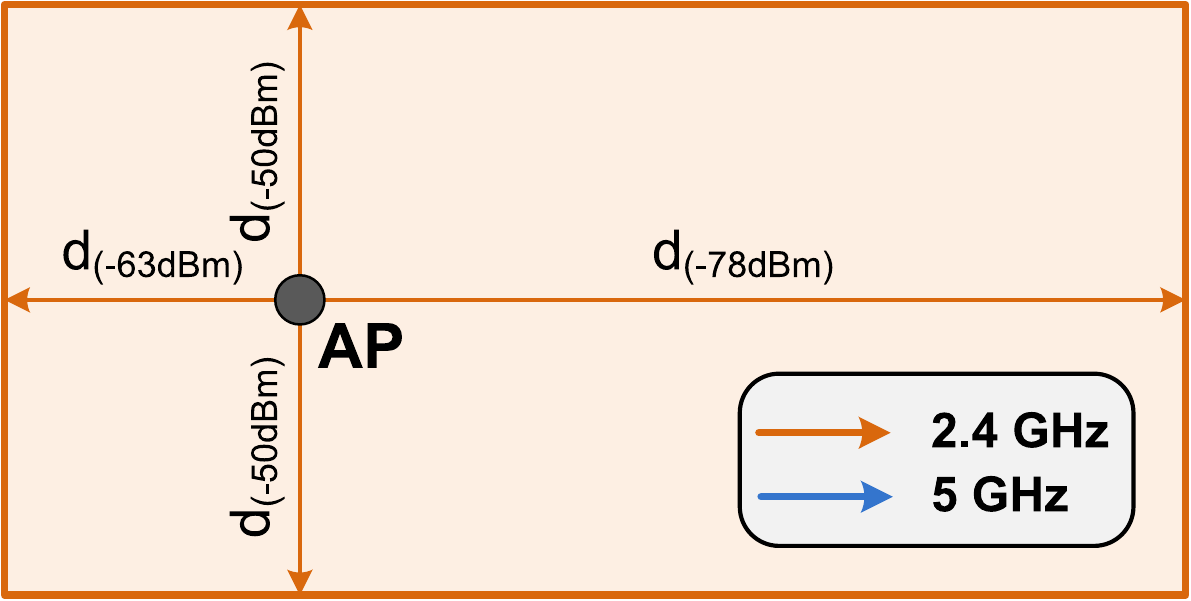}
         \caption{Without Extenders.}
         \vspace{3mm}
       \label{fig:scenario2a}
\end{subfigure}
\begin{subfigure}[]{0.45\textwidth}
       \centering \includegraphics[width=0.75\textwidth]{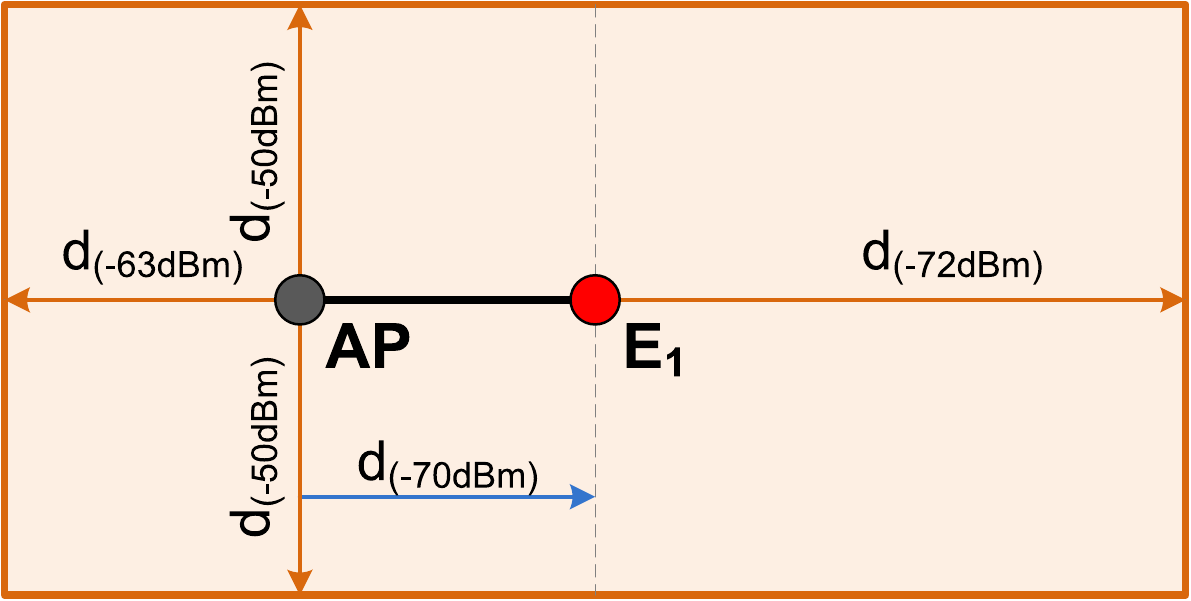}
        \caption{With 1 Extender.}
        \vspace{3mm}
        \label{fig:scenario2b}
\end{subfigure}   
\begin{subfigure}[]{0.45\textwidth}
       \centering \includegraphics[width=0.75\textwidth]{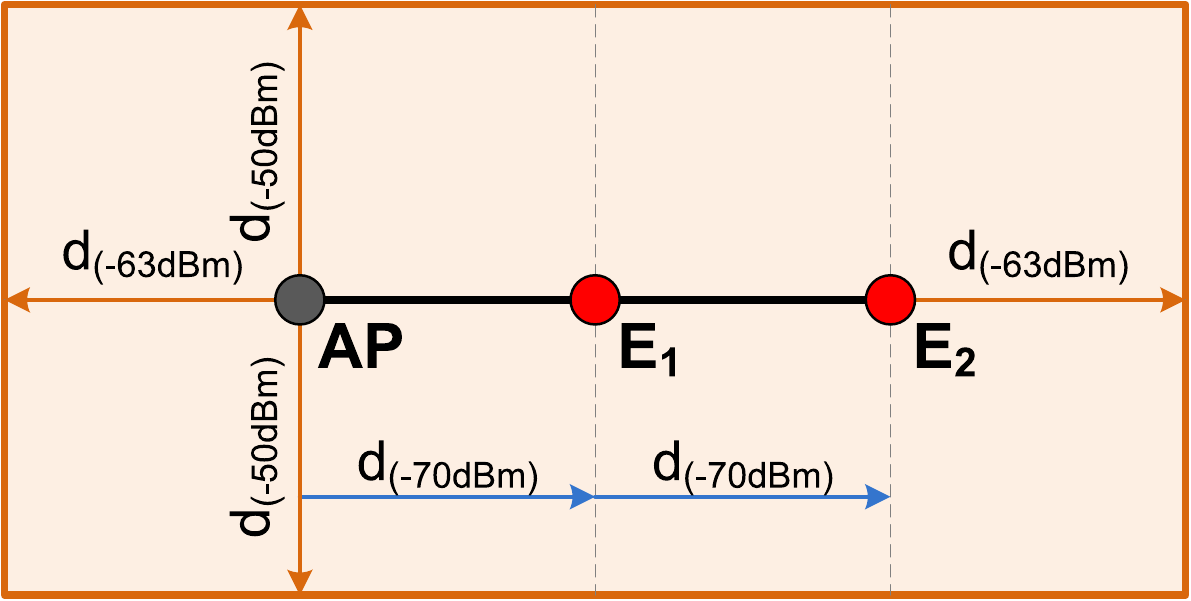}
        \caption{With 2 linked Extenders.}
        \vspace{3mm}
        \label{fig:scenario2c}
\end{subfigure} 

\caption{Network topologies of Scenario \#2.}
\label{fig:scenario2}
\end{figure}

In this case, STAs were deployed within a rectangular area emulating a typical Home WiFi scenario defined according to a set of RSSI values (see Figure~\ref{fig:scenario2}). Three network topologies were there considered: only a single AP, an AP connected to a single Extender, and an AP connected to two linked Extenders. 
    
\subsubsection{Test 2.1: Use of linked Extenders}

To evaluate the effect of linking two Extenders in the backhaul, a set of $N_{\text{STA}} = $ 10 STAs were randomly placed $k =$ 1000 times on all topologies from Figure \ref{fig:scenario2}, with $B_{\text{STA}}$ ranging from $12$ kbps to  $6$ Mbps (i.e., $B_T$ took values from $0.12$ Mbps to $60$ Mbps). Both the \textit{RSSI-based} and the \textit{channel load aware} AP/Extender selection mechanisms were considered (the latter with $\alpha =$ 0.5 to balance access and backhaul links).

\begin{figure*}[th!]
    \centering
    \includegraphics[width=0.75\textwidth]{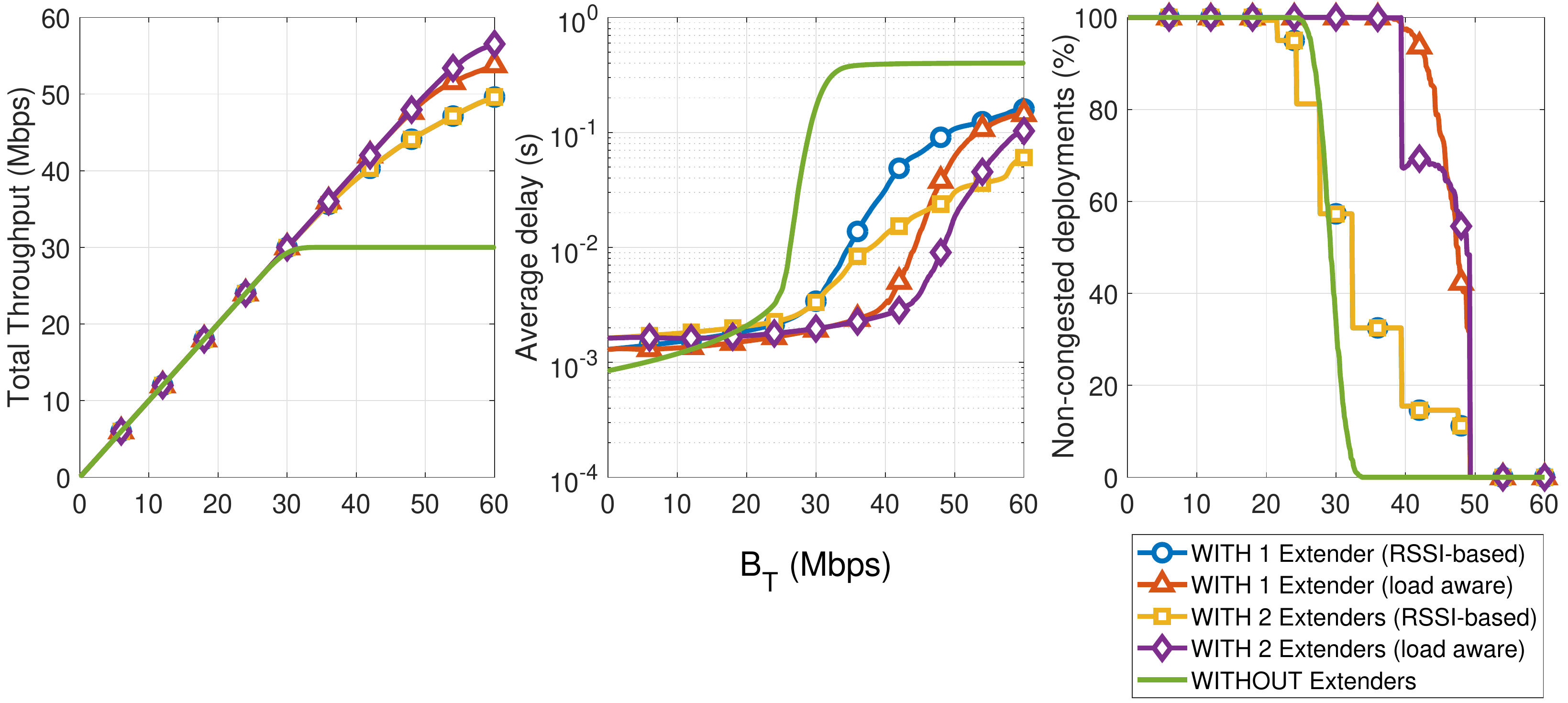}
    \caption{Test 2.1. Use of linked Extenders (multi-channel case).}
    \label{fig:2_1_mix_multi}
\end{figure*}

This test was first performed in a multi-channel case, where the \textit{channel load aware} mechanism was able to avoid network congestion until almost $B_T = 40$ Mbps and improve the performance offered by the \textit{RSSI-based} mechanism, as seen in Figure \ref{fig:2_1_mix_multi}. Furthermore, the use of a second Extender linked to the first one was justified to increase the network's operational range, as shown in Table \ref{tab:results_2_1}.

As for the single channel case, the use of a second Extender (whether under the \textit{RSSI-based} or the \textit{channel load aware} mechanism) here did not result in a significant improvement of any analyzed performance metric. The fact that all STAs (even some of them with low transmission rates) ended up competing for the same channel resources increased the overall occupation and led to congestion for $B_T < 25$ Mbps regardless the number of Extenders.

%\textcolor{red}{[Ja hem vist abans que single channel no acaba d'anar bé. Potser podem eliminar aquest troç i la Figura 8. Bàsicament, els resultats és que no hi ha guany fent servir extenders en res. Això de fet, es pot dir al text amb una frase, però no cal posar la gràfica.]} \textcolor{blue}{[OK, fora Figura 8 i fora el seguent paragraf. Reescrita la idea en el paragraf anterior.] \sout{However, in the single channel case, the \textit{channel load aware} mechanism was not the best option (see Figure \ref{fig:2_1_mix_single}). As already described, the sequential application of its load balancing scheme (i.e., each STA is associated after the other) made STAs not always have the closest AP/Extender as their parent. This fact might be particularly harmful in single channel cases, as the contention for the medium by STAs with low transmission rates increases the overall occupation and may lead to congestion.}}

% \begin{figure*}[h!]
%     \centering
%     \includegraphics[width=0.75\textwidth]{img/2_1_mix_single-eps-converted-to.pdf}
%     \caption{Test 2.1.  Use of linked Extenders (single channel).}
%     \label{fig:2_1_mix_single}
% \end{figure*}

\begin{table*}[h!!!!]
\centering
\footnotesize
\caption{Test 2.1. Use of linked Extenders (network's operational range expressed in terms of $B_T$).}
\label{tab:results_2_1}
\begin{tabular}{|l|l|c|c|c|}
\hline
\multicolumn{2}{|l|}{\multirow{2}{*}{\textbf{Network configuration}}}                                                       & \multicolumn{3}{c|}{$B_T$ (Mbps)}                                                                                                                                                                                  \\ \cline{3-5} 
\multicolumn{2}{|l|}{}                                                                                                      & \textbf{Throughput} $\geqslant$ \textbf{99\%} & \textbf{Delay} $\leqslant$ \textbf{10 ms} & \begin{tabular}[c]{@{}c@{}}\textbf{No congested}\\ \textbf{deployments}\end{tabular} \\ \hline
\multirow{4}{*}{\begin{tabular}[c]{@{}l@{}}\textbf{Multi-}\\ \textbf{channel}\end{tabular}} & \textbf{WITH 2 Extenders (RSSI-based)} & [0, 35.88]                                                                                                     & [0, 38.04]                             & [0, 19.44]                                                                                       \\ \cline{2-5} 
                                                                                   & \textbf{WITH 1 Extenders (RSSI-based)} & [0, 35.88]                                                                                                     & [0, 34.68]                             & [0, 19.44]                                                                                       \\ \cline{2-5} 
                                                                                   & \textbf{WITH 2 Extenders (load aware)} & [0, 53.64]                                                                                                     & [0, 48.36]                             & [0, 32.40]                                                                                       \\ \cline{2-5} 
                                                                                   & \textbf{WITH 1 Extenders (load aware)} & [0, 48.96]                                                                                                     & [0, 44.28]                             & [0, 37.44]                                                                                       \\ \hline
\multirow{4}{*}{\begin{tabular}[c]{@{}l@{}}\textbf{Single}\\ \textbf{channel}\end{tabular}} & \textbf{WITH 2 Extenders (RSSI-based)} & [0, 34.80]                                                                                                     & [0, 32.64]                             & [0, 19.44]                                                                                       \\ \cline{2-5} 
                                                                                   & \textbf{WITH 1 Extenders (RSSI-based)} & [0, 33.72]                                                                                                     & [0, 31.08]                             & [0, 19.44]                                                                                      \\ \cline{2-5} 
                                                                                   & \textbf{WITH 2 Extenders (load aware)} & [0, 30.60]                                                                                                     & [0, 27.84]                             & [0, 25.20]                                                                                       \\ \cline{2-5} 
                                                                                   & \textbf{WITH 1 Extenders (load aware)} & [0, 29.88]                                                                                                     & [0, 27.12]                             & [0, 24.12]                                                                                       \\ \hline
\multicolumn{2}{|l|}{\textbf{WITHOUT Extenders}}                                                                            & [0, 28.80]                                                                                                     & [0, 26.28]                             & [0, 23.76]                                                                                       \\ \hline
\end{tabular}
\end{table*}

\subsubsection{Test 2.2: Impact of access and backhaul links}

Assuming the network topology from Figure \ref{fig:scenario2c} with 2 linked Extenders, the effect of $\alpha$ parameter on the \textit{channel load aware} AP/Extender selection mechanism was studied for $\alpha = \left\lbrace 0, 0.25, 0.5, 0.75, 1 \right\rbrace$ and $B_{\text{STA}} = \left\lbrace 1.8, 3, 4.2, 5.4 \right\rbrace$ Mbps (i.e., a total network traffic of $B_T = \left\lbrace 18, 30, 42, 54 \right\rbrace$ Mbps, respectively).

\begin{figure*}[th]
\centering
\begin{subfigure}[t]{0.4\textwidth}
    \centering
    \includegraphics[width=\textwidth]{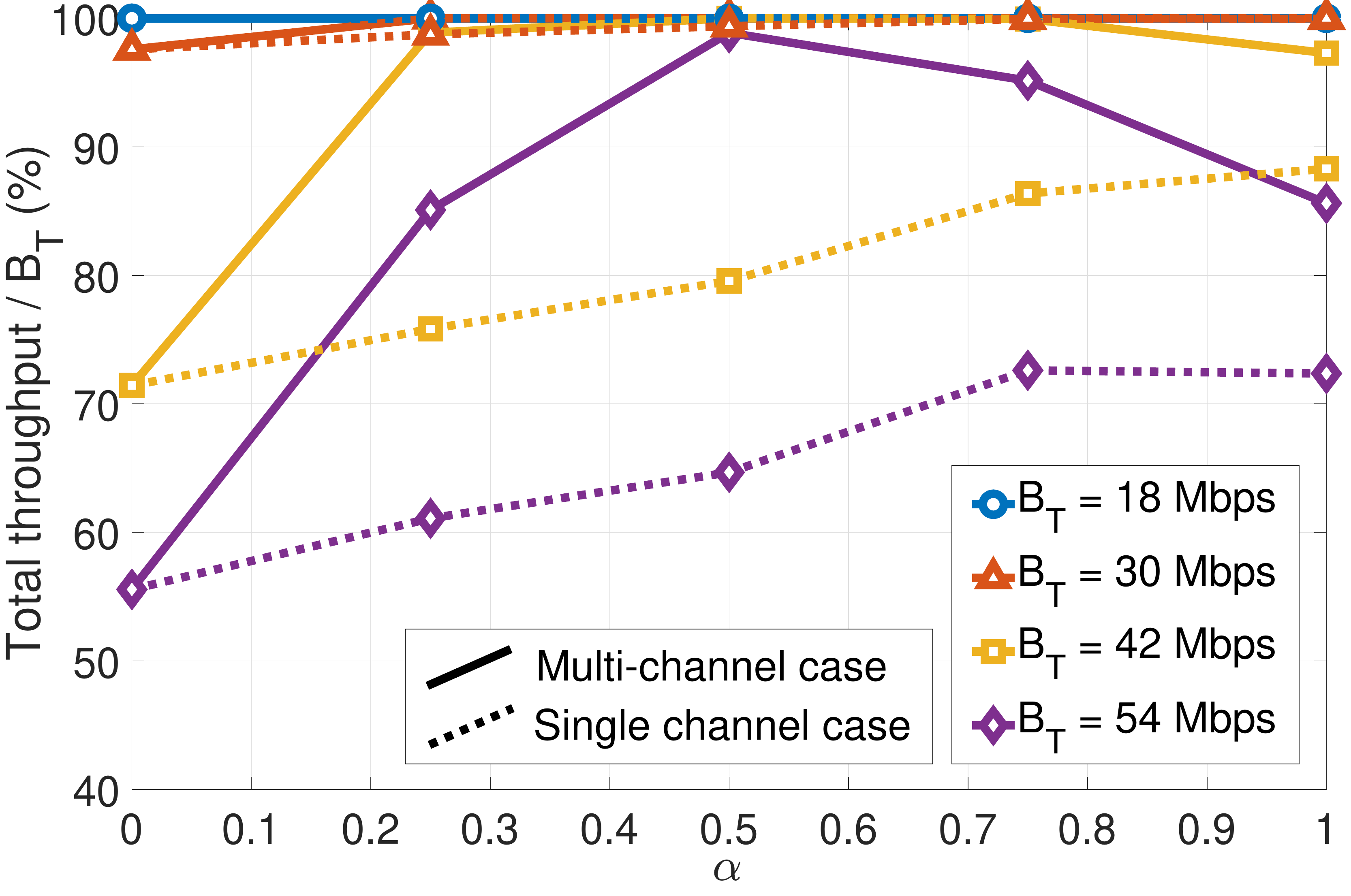}
    \caption{Percentage of throughput.}
    \label{fig:2_2_throughput}
\end{subfigure}
\begin{subfigure}[t]{0.4\textwidth}
    \centering \includegraphics[width=\textwidth]{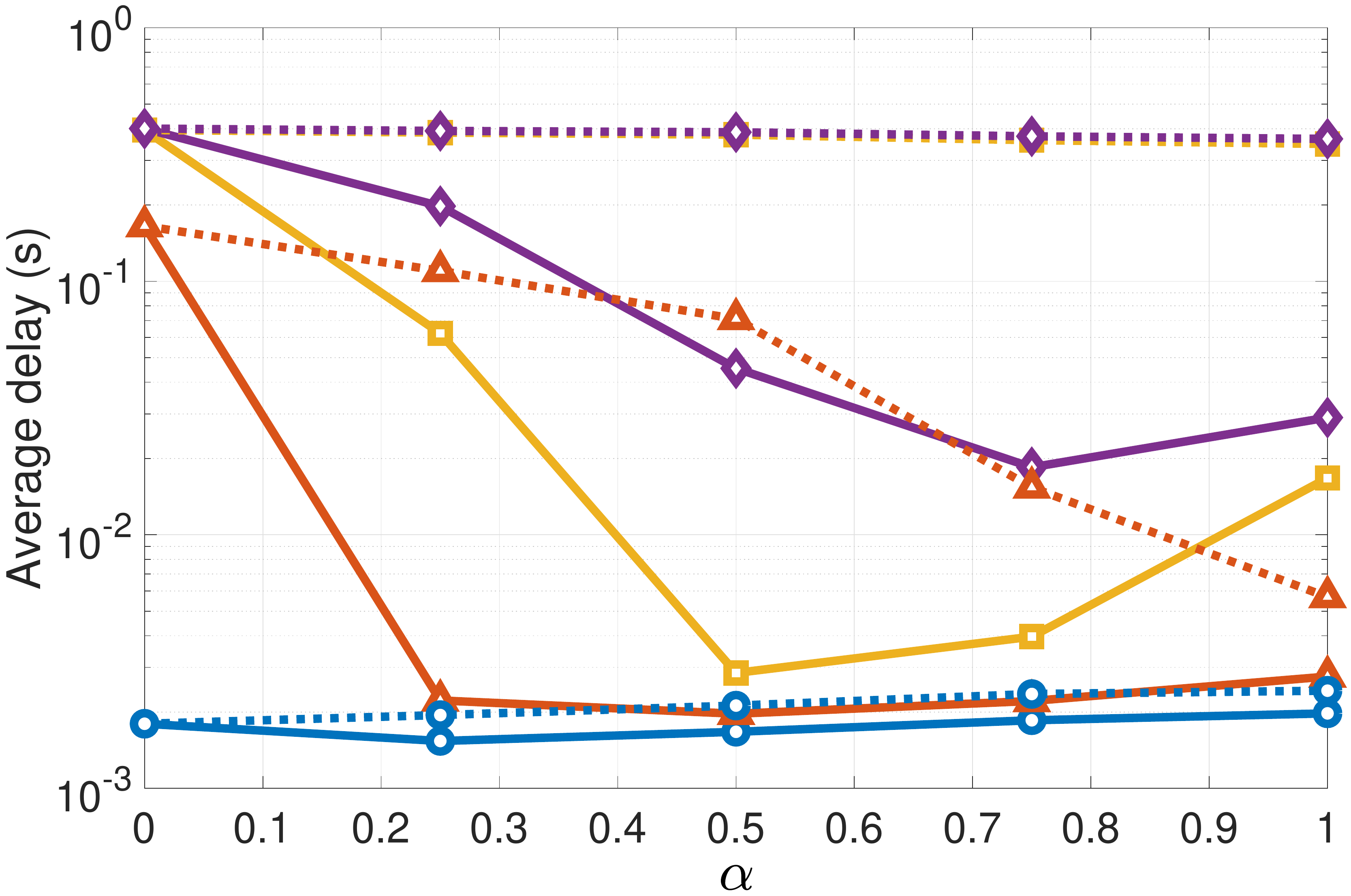}
    \caption{Average delay.}
    \label{fig:2_2_delay_avg}
\end{subfigure} 
\caption{Test 2.2. Impact of access and backhaul links.}
\label{fig:2_2}
\end{figure*}

As shown in Figure \ref{fig:2_2}, values of $\alpha \in [0.5, 0.75]$ in the multi-channel case were able to guarantee the best network performance in terms of throughput ($> 95\%$) and delay ($< 50$ ms) for any considered $B_T$ value. In fact, to give all the weight in (\ref{eq:score}) either to the access link ($\alpha~=$~1) or to the backhaul links ($\alpha~=$~0) never resulted in the best exploitation of network resources.

On the other hand, the best performance in the single channel case was achieved when $\alpha =$ 1; that is, when the \textit{channel load aware} mechanism behaved as the \textit{RSSI-based} one and therefore only the RSSI value was taken into account to compute the best AP/Extender for each STA.\footnote{In the single channel case, $C_{a_{i,j}}$ element in (\ref{eq:score}) is the same for any access link. Then, if $\alpha =$ 1 (i.e., all the weight is given to the access link), the decisive factor is $\text{RSSI}_{i,j}$.}

\subsubsection{Test 2.3: Share of IEEE 802.11k/v capable STAs}

The \textit{channel load aware} AP/Extender selection mechanism can be executed by IEEE 802.11k/v capable STAs without detriment to the rest of STAs, which would continue using the \textit{RSSI-based} mechanism as usual. This test intended to evaluate this effect on overall network's performance.

Assuming again the network topology from Figure \ref{fig:scenario2c} with 2 linked Extenders, the effect of the share of IEEE 802.11k/v capable STAs (here noted as $\beta$) on the \textit{channel load aware} mechanism was studied for $\alpha =$ 0.5, $\beta = \left\lbrace 0, 25, 50, 75, 100 \right\rbrace$ \%, and $B_{\text{STA}} = \left\lbrace 1.8, 3, 4.2, 5.4 \right\rbrace$ Mbps (i.e., a total network traffic of $B_T = \left\lbrace 18, 30, 42, 54 \right\rbrace$ Mbps, respectively).

\begin{figure*}[th]
\centering
\begin{subfigure}[t]{0.4\textwidth}
    \centering
    \includegraphics[width=\textwidth]{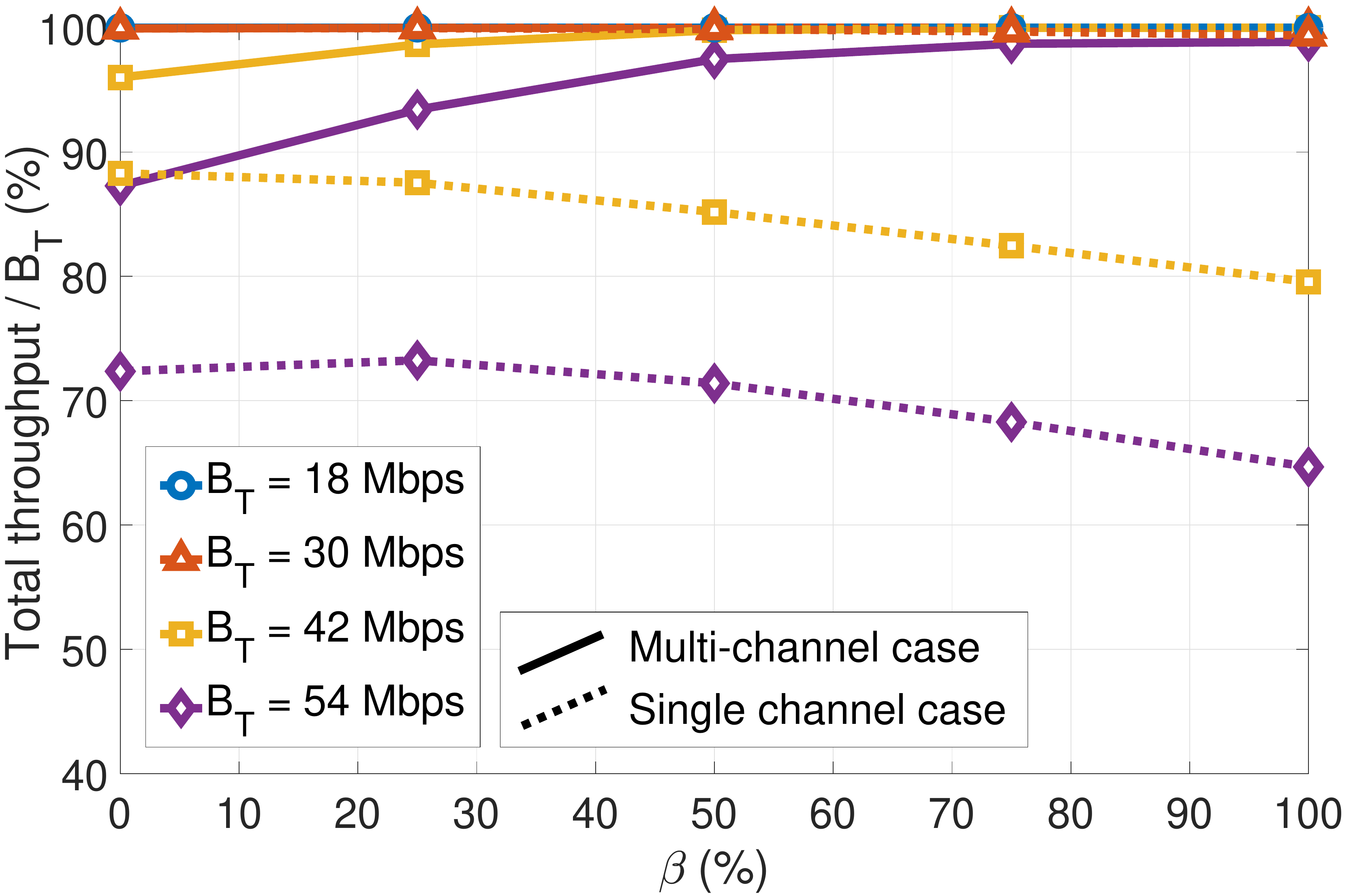}
    \caption{Percentage of throughput.}
    \label{fig:2_3_throughput}
\end{subfigure}
\begin{subfigure}[t]{0.4\textwidth}
    \centering \includegraphics[width=\textwidth]{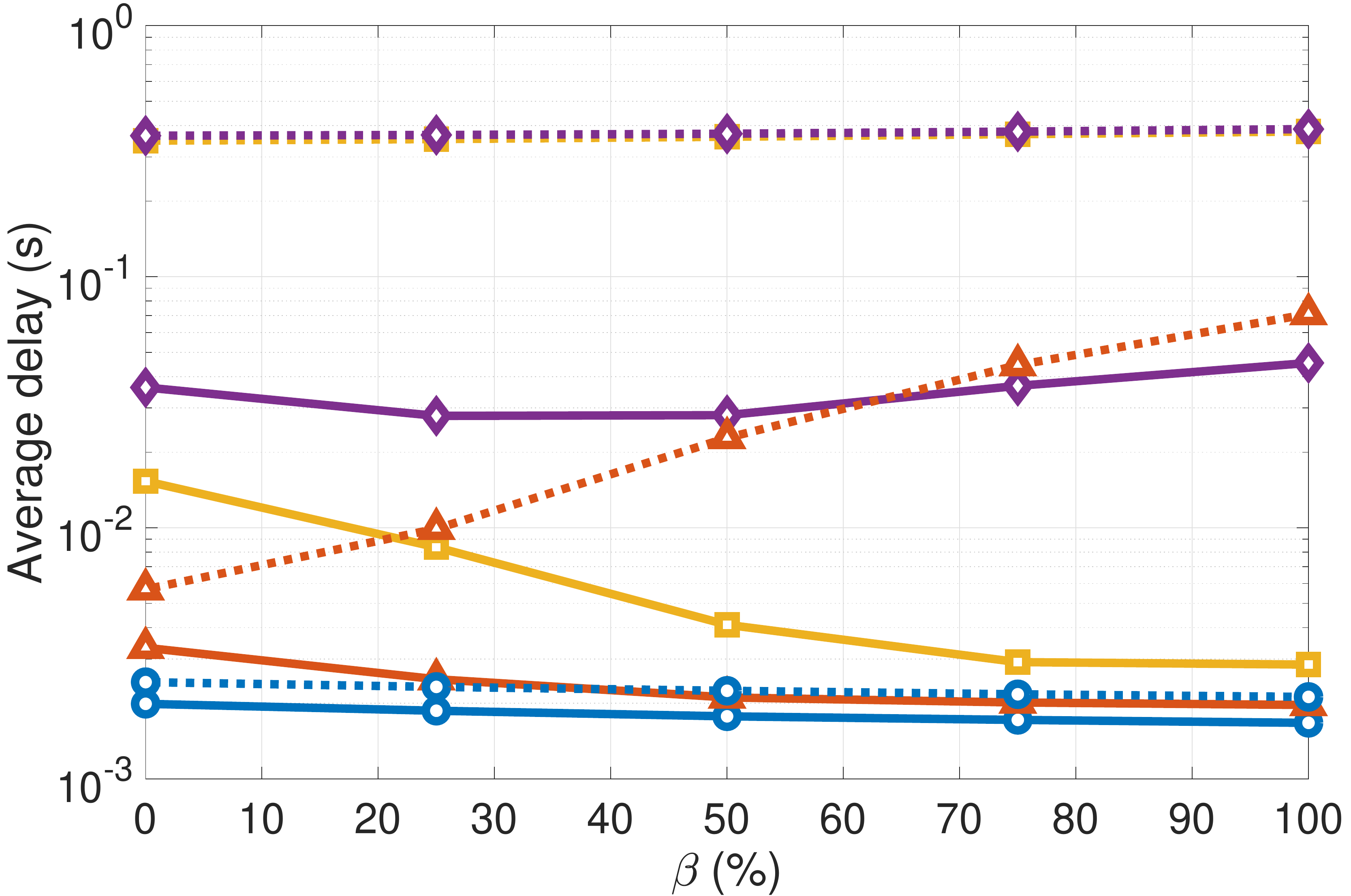}
    \caption{Average delay.}
    \label{fig:2_3_delay_avg}
\end{subfigure} 
\caption{Test 2.3. Share of IEEE 802.11k/v capable STAs.}
\label{fig:2_3}
\end{figure*}

As shown in Figure \ref{fig:2_3}, there was a clear trend in the multi-channel case that made network's performance grew together with the share of IEEE 802.11k/v capable STAs, even ensuring more than 95\% of throughput for any considered $B_T$ value when half or more of STAs were IEEE 802.11k/v capable.

On the contrary, in the single channel case the best results were achieved when $\beta =$ 0 or, in other words, when none STA had IEEE 802.11k/v capabilities and therefore all of them applied the traditional \textit{RSSI-based} mechanism.

% ###########################################################

\subsubsection{Test 2.4: Interference from external networks}

We aimed to evaluate the potential negative effect that the presence of neighboring WLANs could have on the \textit{channel load aware} AP/Extender selection mechanism, and verify if that mechanism continued outperforming the \textit{RSSI-based} one in terms of total throughput and average delay. 

\begin{figure}[th!!!!]
    \centering
    \includegraphics[width=0.45\textwidth]{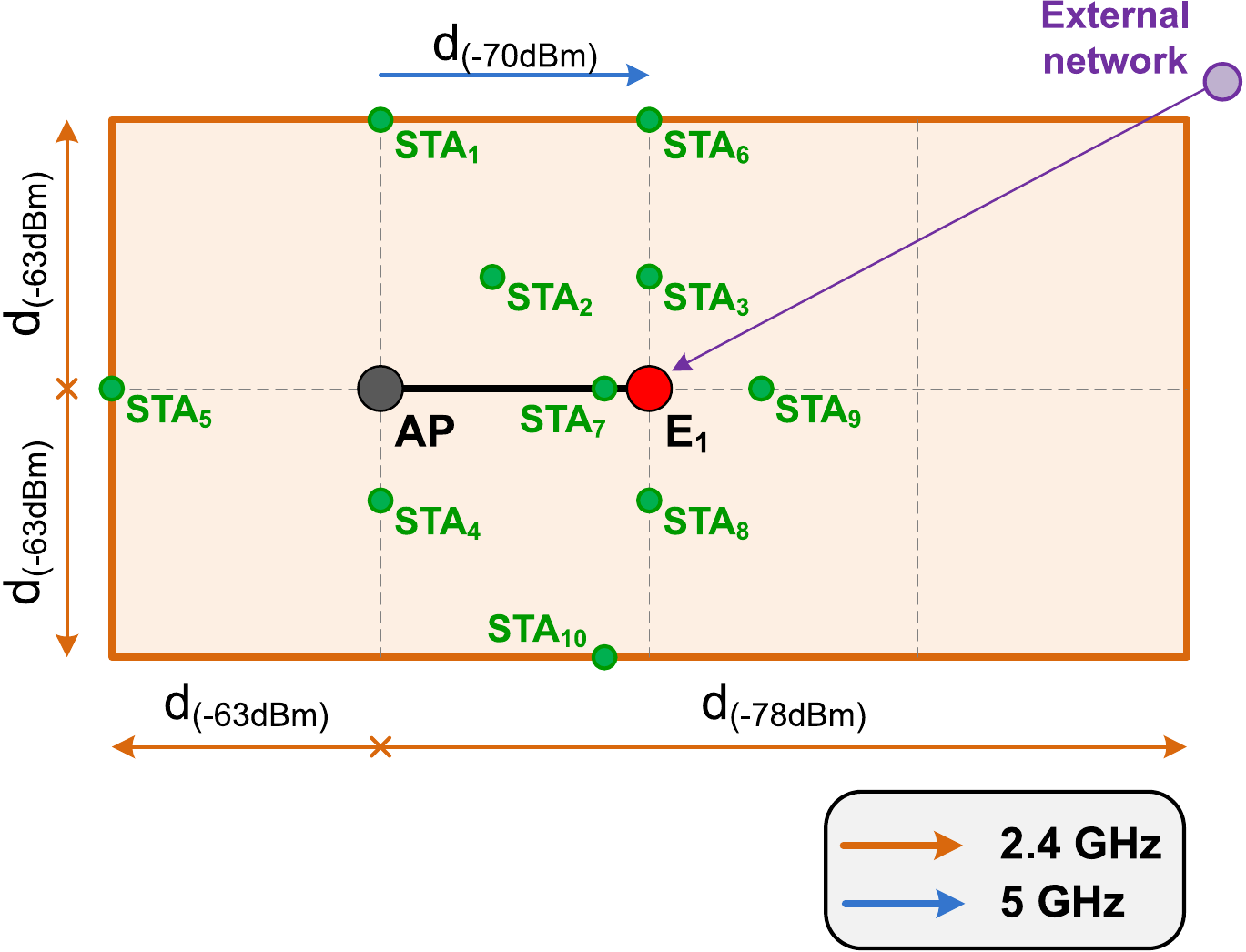}
    \caption{Network topology and STA deployment of Test 2.4.}
    \label{fig:scenario2d}
\end{figure}

A particular scenario with an AP, an Extender and 10 STAs was considered following the deployment shown in Figure \ref{fig:scenario2d}, where the Extender shared its access link channel at 2.4 GHz band with an external network. Whereas the traffic load of each STA was set to $B_{\text{STA}} = 4.32$ Mbps, the load of the external network ranged from $B_{\text{EXT}} = 0$ Mbps to $B_{\text{EXT}} = 12$ Mbps. %Selected values for the $\alpha$ parameter were $\alpha = \left\lbrace 0.5, 0.75, 1 \right\rbrace$.

\begin{figure*}[h]
    \centering
    \begin{subfigure}[t]{0.4\textwidth}
        \centering
        \includegraphics[width=\textwidth]{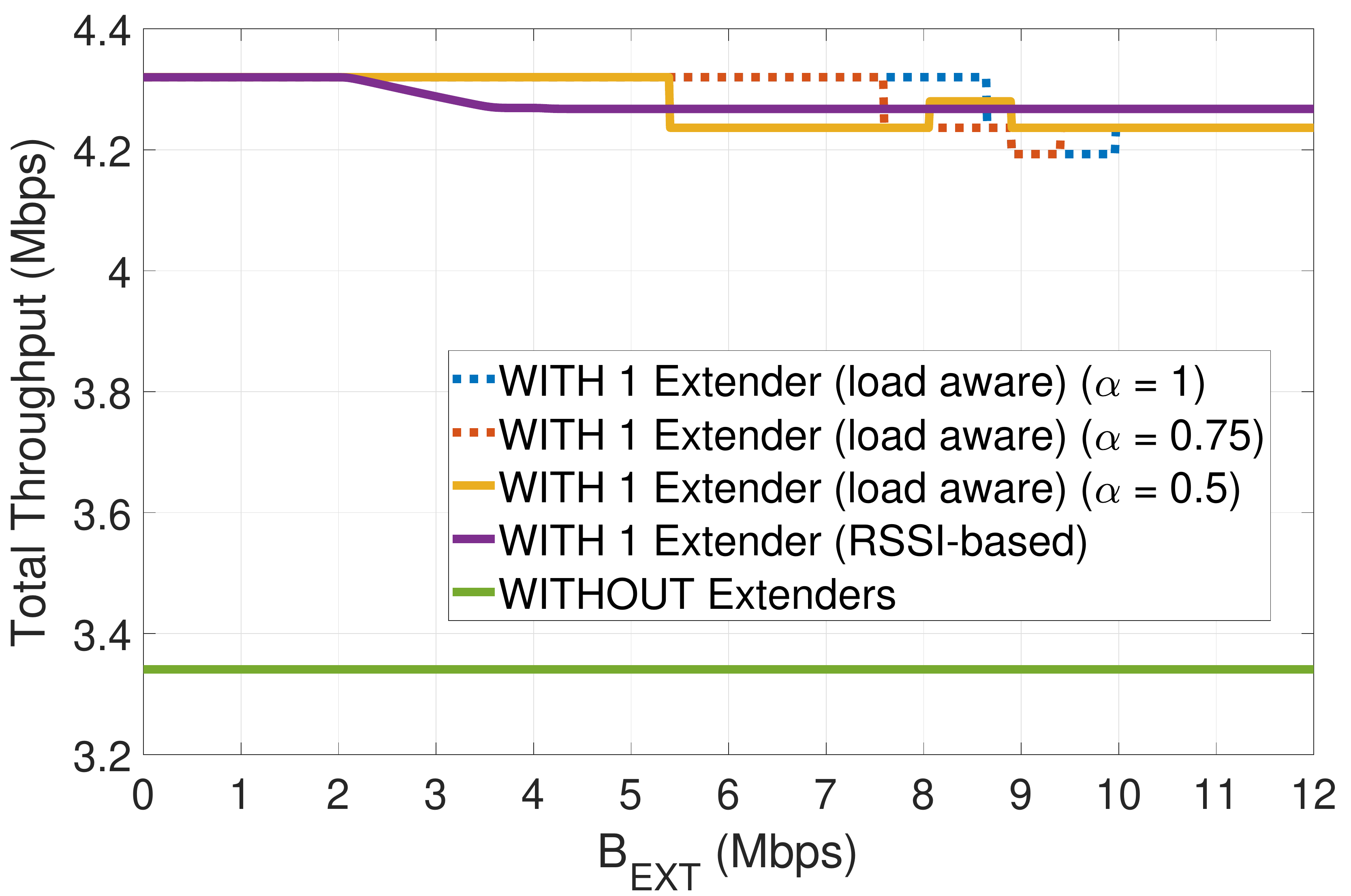}
        \caption{Total throughput.}
        \label{fig:2_4_throughput}
    \end{subfigure}
    \begin{subfigure}[t]{0.4\textwidth}
        \centering \includegraphics[width=\textwidth]{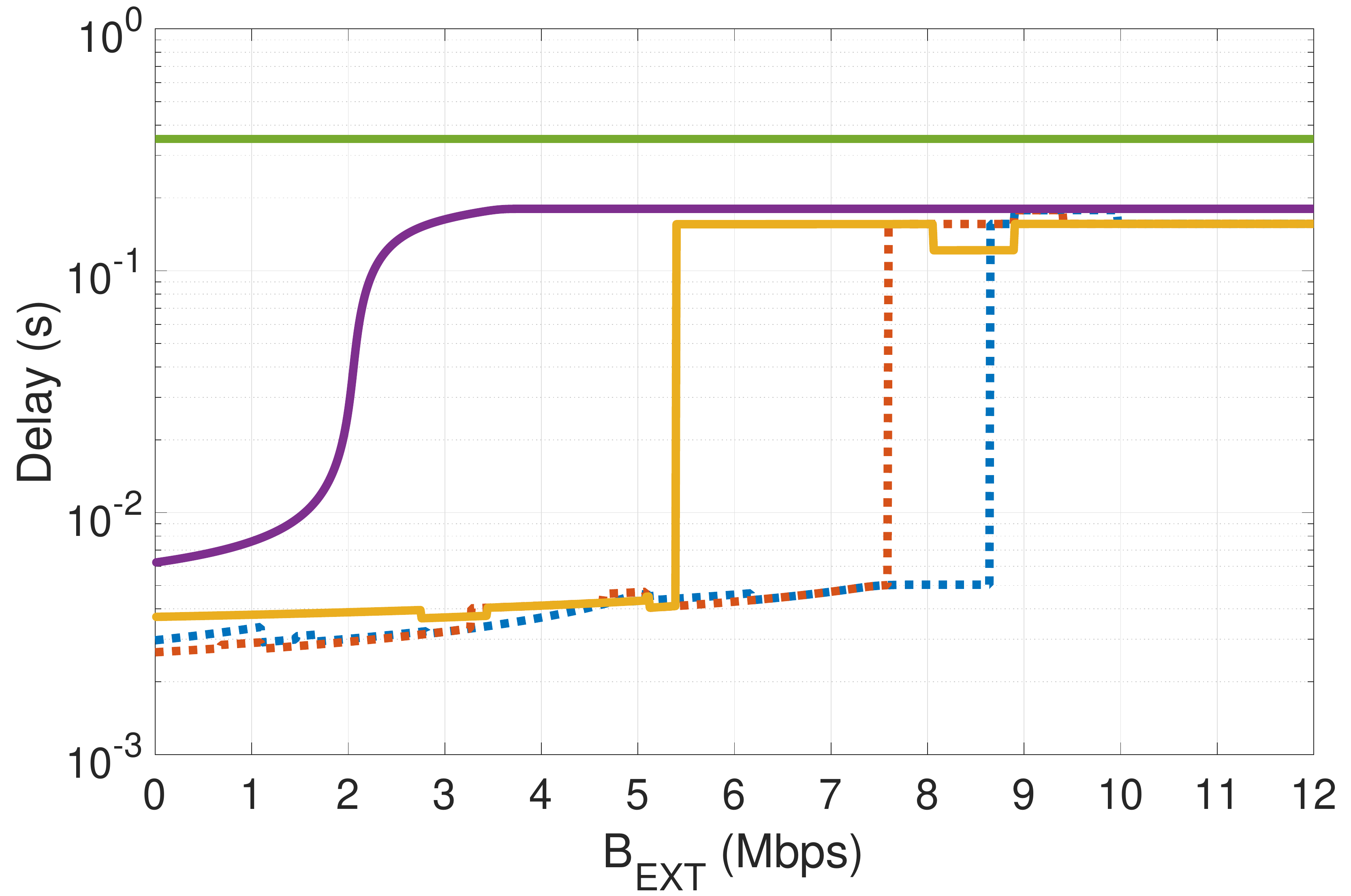}
        \caption{Average delay.}
        \label{fig:2_4_delay_avg}
    \end{subfigure} 
    \caption{Test 2.4. Interference from external networks.}
    \label{fig:2_4}
\end{figure*}

Figure \ref{fig:2_4_throughput} shows that, for any considered $\alpha$ value, the \textit{channel load aware} mechanism was able to deliver $100\%$ of throughput for higher $B_{\text{EXT}}$ values than the \textit{RSSI-based} configuration, having the highest $\alpha$ values the best performance. The topology without Extenders, here maintained as a reference, again demonstrates the utility of Extenders in such Home WiFi scenarios. 

The average delay of STAs followed the same trend (see Figure \ref{fig:2_4_delay_avg}), having again the \textit{channel load aware} mechanism the best performance, maintaining it below $5$ ms in any configuration given $B_{\text{EXT}} < 5$ Mbps. Observing the delay, it is worth noting the difference between the gradual delay increase in the \textit{RSSI-based} mechanism (due to the progressive saturation of the access link to the Extender when $B_{\text{EXT}} \in [1.5,3.5]$ Mbps) in comparison with its abrupt change in the \textit{channel load aware} one. This was due to a different AP/Extender selection of one or more STAs from a given $B_{\text{EXT}}$ value on.

% ----------------------------------------------------------------------------
% ----------------------------------------------------------------------------
% ----------------------------------------------------------------------------
% ----------------------------------------------------------------------------

\section{Performance of the AP/Extender selection mechanism in a real deployment}
\label{deployment}

A testbed was deployed at Universitat Pompeu Fabra (UPF) to emulate a Home WiFi network and, therefore, further study the benefits of using Extenders and the performance of the \textit{channel load aware} AP/Extender selection mechanism.

The hardware employed consisted of an AP, an Extender, and 5 laptops acting as traffic generation STAs. A sixth laptop was connected to the AP through Ethernet to act as the traffic sink. The AP and the Extender were placed at a distance that guaranteed $\text{RSSI}_{\text{AP,E}}$ = -70 dBm at 5 GHz, as in the previous simulated scenarios. As for the 2.4 GHz band, non-overlapping communications were ensured by using orthogonal channels.

\begin{figure}[h!]
    \centering
    \begin{subfigure}[t]{0.2\textwidth}
        \centering \includegraphics[width=\textwidth]{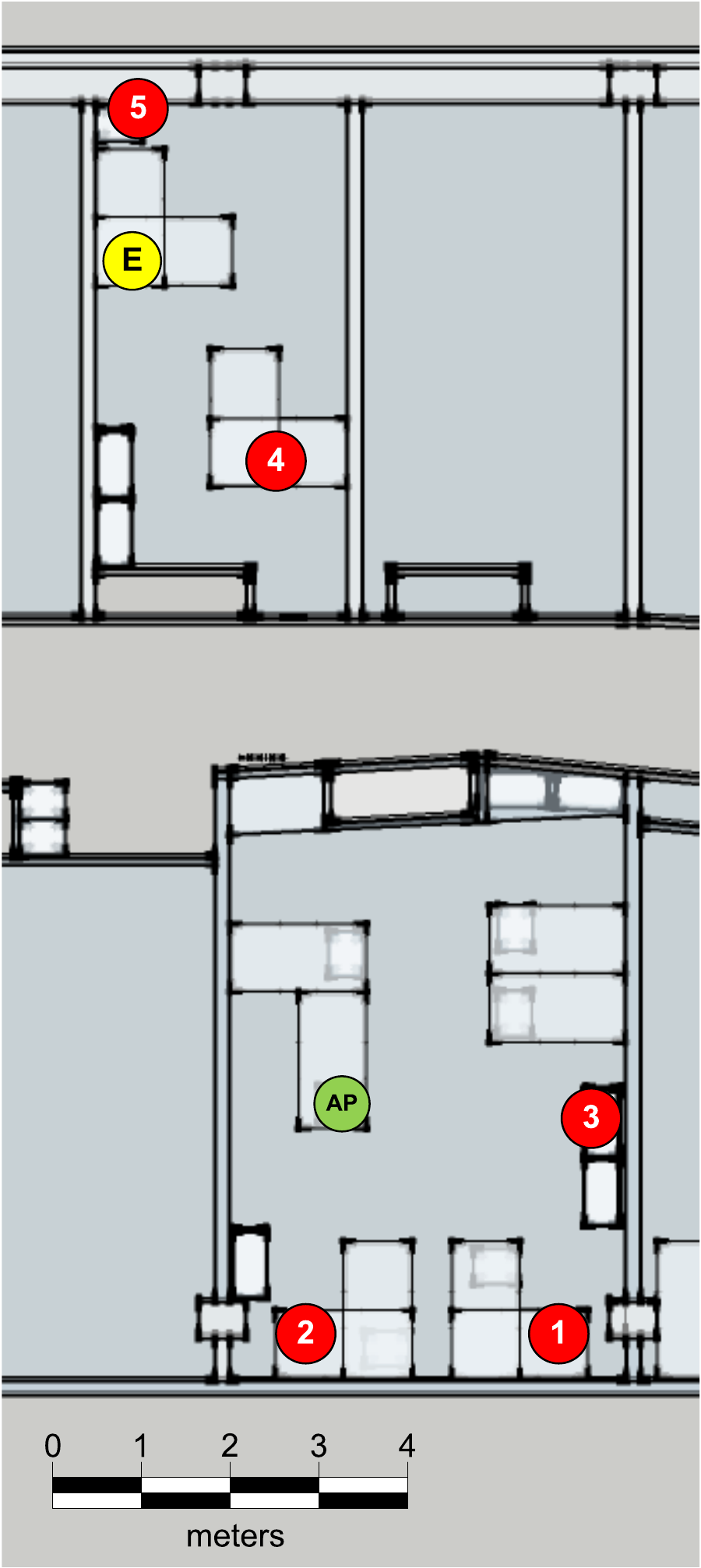}
        \caption{Testbed \#1.}
        \label{fig:testbed1}
    \end{subfigure} 
    \begin{subfigure}[t]{0.2\textwidth}
        \centering \includegraphics[width=\textwidth]{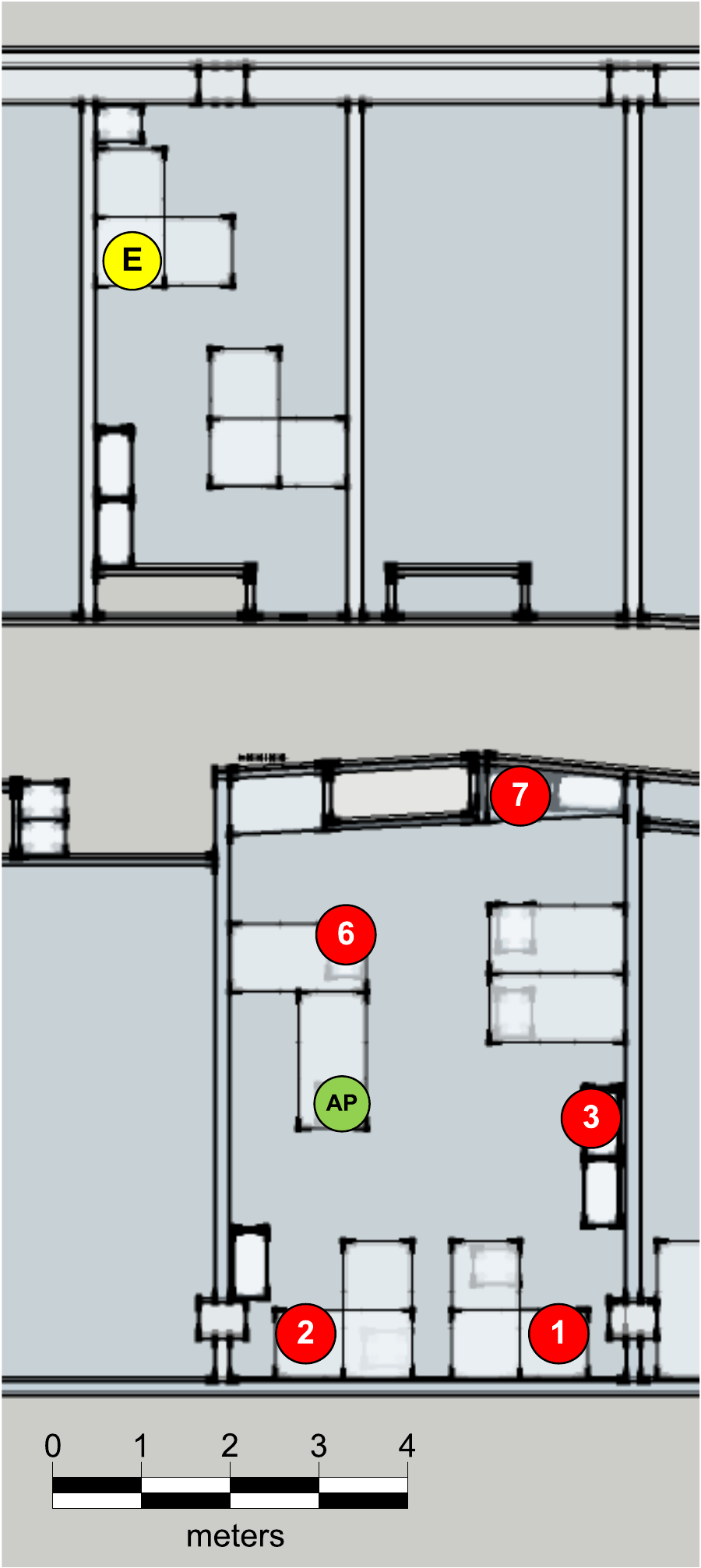}
        \caption{Testbed \#2.}
        \label{fig:testbed2}
    \end{subfigure} 
    \caption{Plan map of testbeds performed at UPF and placement of network devices.}
    \label{fig:testbeds}
\end{figure}

STAs were deployed in 2 different sets of positions (see Figure \ref{fig:testbeds}). Then, using the RSSI and load parameters from each STA, all network links were obtained according to the appropriate AP/Extender selection mechanism. These links were then set in the real deployment to get the performance results.

Tests were performed using \textit{iPerf}\footnote{iPerf main website: \url{https://iperf.fr/}} version 2.09 or higher, which allowed the use of enhanced reports that included both the average throughput and the delay of the different network links. The clocks of the STAs needed to be synchronized for the delay calculation, and this was achieved using the network time protocol (NTP)\footnote{NTP main website: \url{http://www.ntp.org/}}.

UDP traffic was used in all \textit{iPerf} tests. Several traffic loads were used in each test, and 5 trials were performed for each traffic load. Each trial lasted 60 seconds. Clocks were re-synchronized before every new load was tested (i.e., every 5 trials), leading to an average clock offset of $+/- 0.154$ ms. All trials were performed during non-working ours, and there were no other WiFi users at UPF during the tests.  

%All trials were performed during the weekend to ensure channels were clear of interference. 

\begin{table*}[]
\centering
\footnotesize
\caption{RSSI values received by STAs from AP/Extender and selected next hop in Testbed \#1 and \#2.}
\label{tab:routing}
\begin{tabular}{|c||c|c||c|c||c|c|c|c|c|c|c|c|}
\hline
\multirow{4}{*}{} & \multicolumn{2}{c||}{\multirow{2}{*}{\begin{tabular}[c]{@{}c@{}}\textbf{RSSI}\\ \textbf{(dBm)}\end{tabular}}} & \multicolumn{2}{c||}{\textbf{Testbed \#1}}                                   & \multicolumn{6}{c|}{\textbf{Testbed \#2}}                                                                                                                                                  \\ \cline{4-11} 
                  & \multicolumn{2}{c||}{}                                                                               & \textbf{Only AP}                     & \textbf{AP + Extender}                                   & \multicolumn{6}{c|}{\textbf{AP + Extender}}                                                                                                         \\ \cline{2-11} 
                  & \multirow{2}{*}{\textbf{AP}}                      & \multirow{2}{*}{\textbf{E}}                     & \multirow{2}{*}{\textbf{RSSI-based}} & \multirow{2}{*}{\textbf{RSSI-based}}  & \multirow{2}{*}{\textbf{RSSI-based}} & \multicolumn{5}{c|}{\begin{tabular}[c]{@{}c@{}}\textbf{Load aware}\\ \textbf{(in function of $B_T$ in Mbps)}\end{tabular}} \\ \cline{7-11} 
                  &                                                   &                                                 &                                      &                                                                            &                                      & \textbf{5}           & \textbf{37.5}      & \textbf{50}      & \textbf{75}     & \textbf{100}     \\ \hline
\textbf{STA \#1}      & -43                                               & -66                                             & AP                                   & AP                                                                     & AP                                   & AP                            & AP                & AP               & AP              & AP              \\ \hline
\textbf{STA \#2}      & -31                                               & -69                                             & AP                                   & AP                                                                      & AP                                   & AP                           & AP                & AP               & AP              & E               \\ \hline
\textbf{STA \#3}      & -38                                               & -67                                             & AP                                   & AP                                                                     & AP                                   & AP                           & E                 & E                & E               & AP              \\ \hline
\textbf{STA \#4}      & -59                                               & -41                                             & AP                                   & E                                                                        & -                                    & -                             & -                 & -                & -               & -               \\ \hline
\textbf{STA \#5}      & -65                                               & -35                                             & AP                                   & E                                                                      & -                                    & -                              & -                 & -                & -               & -               \\ \hline\hline
\textbf{STA \#6}      & -41                                               & -51                                             & -                                    & -                                                                      & AP                                   & AP                            & AP                & AP               & AP              & AP              \\ \hline
\textbf{STA \#7}      & -46                                               & -52                                             & -                                    & -                                                                      & AP                                   & E                              & E                 & E                & E               & AP              \\ \hline
\end{tabular}
\end{table*}

% --------------------------
% --------------------------

\subsection{Experiment 1: On the benefits of using Extenders}

Testbed \#1 was designed to analyze the performance of a network that consisted of one AP and one Extender, considering only the \textit{RSSI-based} association mechanism. The device placement for this experiment can be found in Figure \ref{fig:testbed1}. Two cases were considered: the first one was the deployment without the Extender, meaning that all STAs were forced to associate to the AP. The second case did consider the Extender, allowing STAs to associate to either the AP or the Extender. The association for each case can be found in Table \ref{tab:routing}, as well as the RSSI of each STA for both the AP and the Extender. 

In the first case, where all STAs associated to the AP, we can observe that the RSSI was very low for STAs \#4 and \#5, as expected. Once we added the Extender in the second case, STAs \#4 and \#5 were associated to it, and so they improved their RSSI. Specifically, STA \#4 got an increase of 30.51\%, and STA \#5 experienced an increase of 46.15\%, respectively. The average RSSI of the different links was also increased, going from -47.20 dBm to -37.60 dBm (i.e., 20.34\% higher).

Three different total network traffic loads ($B_T$), as a result of the corresponding traffic load per STA ($B_{\text{STA}}$), were tested in each case, starting with $B_{\text{STA}} = 1$ Mbps (i.e., $B_T = 5$ Mbps), then $B_{\text{STA}} = 3$ Mbps (i.e., $B_T = 15$ Mbps), and lastly $B_{\text{STA}} = 7.5$ Mbps (i.e., $B_T = 37.5$ Mbps).

\begin{figure}[t!!]
    \centering
    \includegraphics[width=0.48\textwidth]{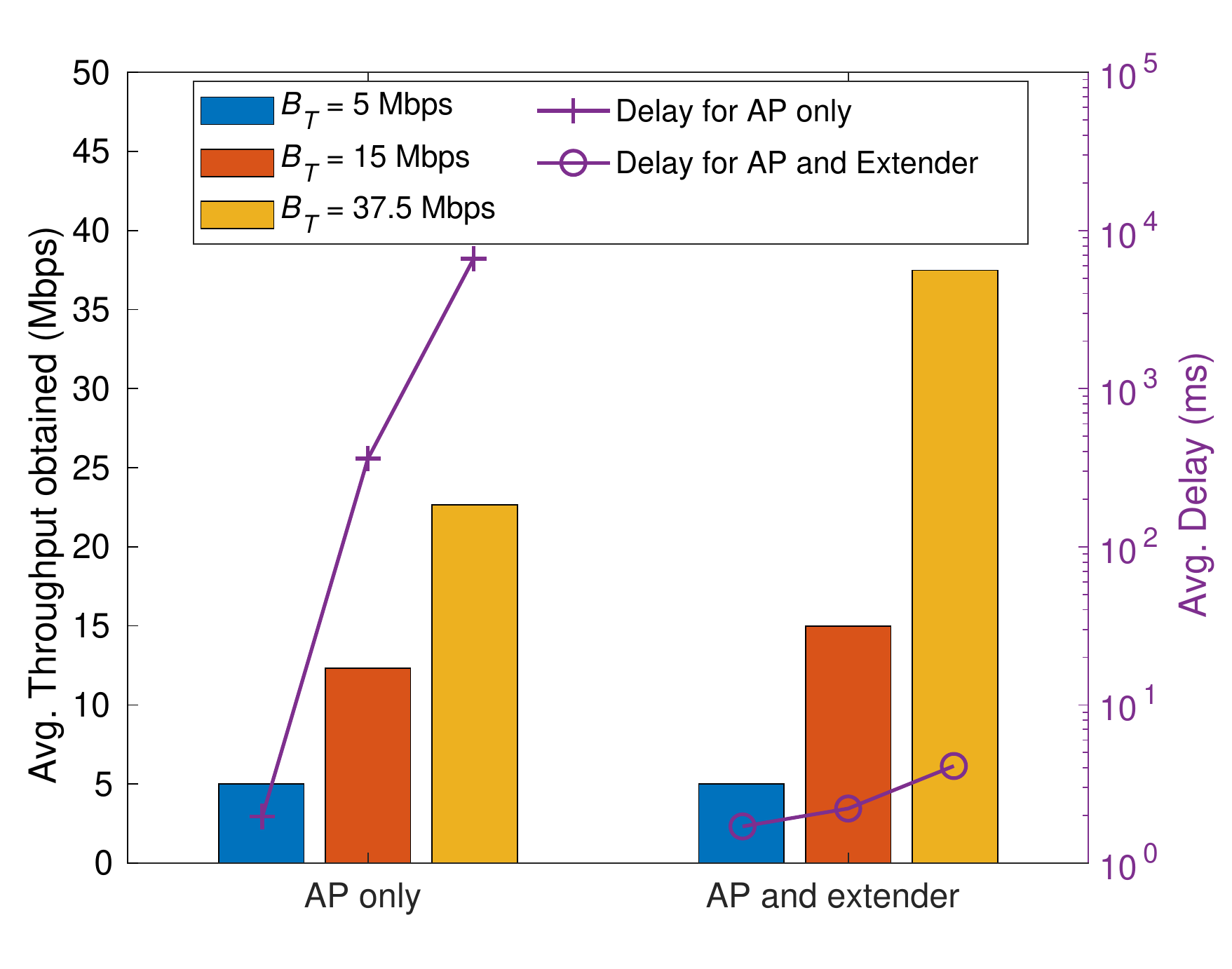}
    \caption{Throughput and delay achieved in Testbed \#1.}
    \label{s1brief}
\end{figure}

Figure \ref{s1brief} shows the throughput achieved for each load, as well as the average delay for the network. Regardless the presence of the Extender, 100\% of throughput was achieved for $B_T = 5$ Mbps. Higher differences appeared for $B_T = 15$ Mbps and $B_T = 37.5$ Mbps, as without the Extender the network was saturated, whereas 100\% of the desired throughput was achieved when using the Extender. 

\begin{figure}[ht]
    \centering
    \includegraphics[width=0.48\textwidth]{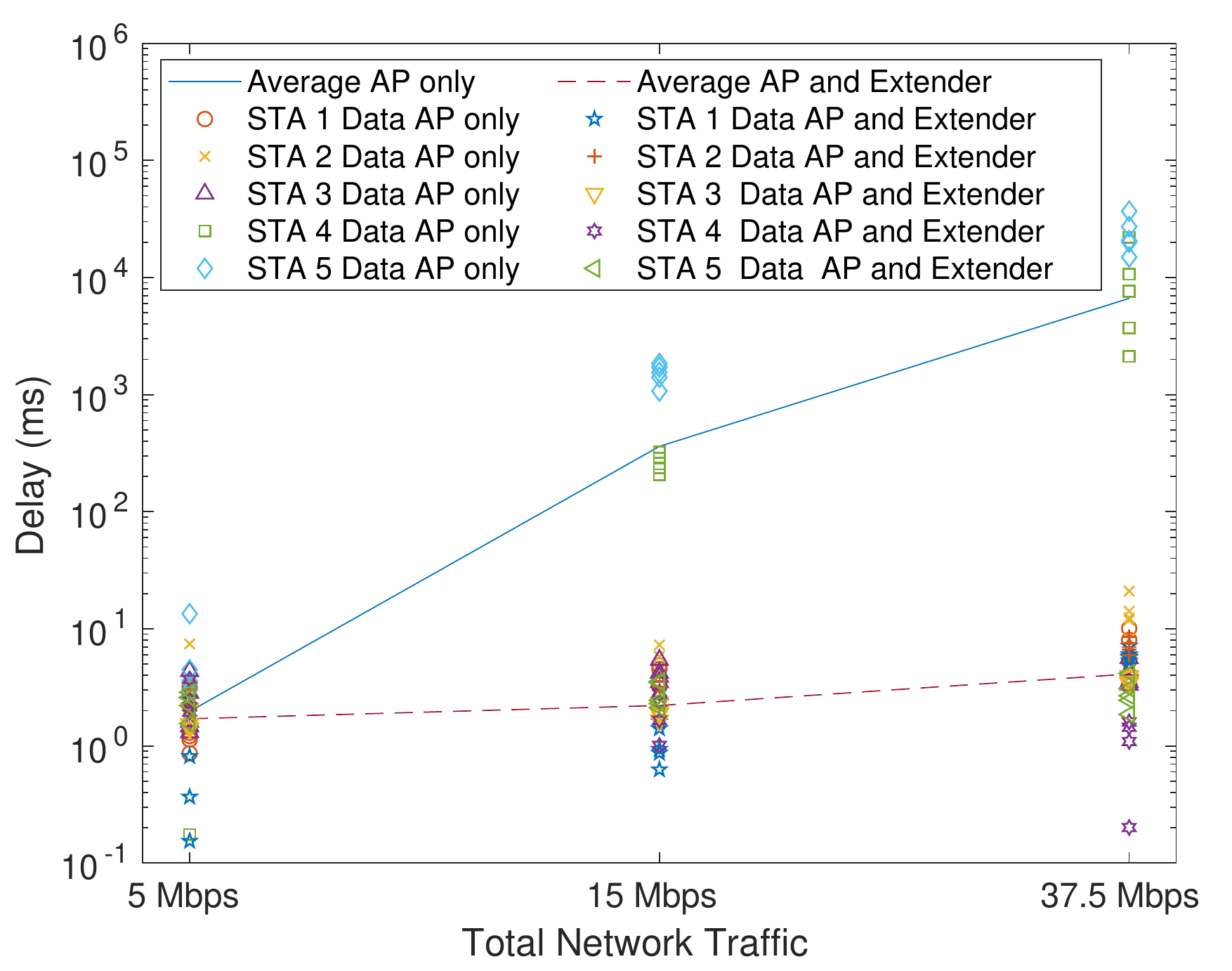}
    \caption{Average delay by STA in Testbed \#1.}
    \label{s1STADelay}
\end{figure}

The use of an Extender is also beneficial for the average delay, as even in the worst case, when $B_T = 37.5$ Mbps, this value was reduced from 6633.84 ms to 4.10 ms. The reason of such huge delays when not using Extenders can be observed in Figure \ref{s1STADelay}, where the delay breakdown per STA shows how STA \#4 and STA \#5 influenced  the overall average values. 

In this experiment we have shown that the use of Extenders in a Home WiFi network can be beneficial beyond the extension of the coverage area, increasing both the minimum and the average RSSI for the whole network, as well as achieving higher throughput capacity and lower delays. These results therefore support our previous simulations, whose results are compiled in Table \ref{tab:association_1_1}, Table \ref{tab:results_1_3}, and Figure \ref{fig:1_3_mix}.

% --------------------------

\subsection{Experiment 2: Validation of the channel load aware AP/Extender selection mechanism}

Testbed \#2 was deployed following Figure \ref{fig:testbed2} to evaluate the performance of the \textit{channel load aware} AP/Extender selection mechanism and compare it to the \textit{RSSI-based} mechanism. The AP and the Extender were always active and in non-overlapping channels. All STAs were inside the office that contained the AP, and we applied both selection mechanisms to every STA. For the \textit{channel load aware} mechanism, the $\alpha$ used was 0.5; i.e., the influence of the access and the backhaul links was the same when selecting an AP/Extender.

Five increasing loads were used to compare the performance of the \textit{RSSI-based} and the \textit{channel load aware} selection mechanisms. The resulting association for all STAs, as well as their traffic loads can be found in Table \ref{tab:routing}, where we can observe that at least one STA was always associated to the Extender when using the \textit{channel load aware} mechanism, thus resulting in better use of network resources.%\textcolor{red}{[On s'observa això? Aquets comentari té sentit si es parla de varies repeticions del mateix, i en totes, almenys un client passa a l'extender... pero en altres casos, pot no ser correcte.]} \textcolor{Plum}{Parlo nomes dels tests eh! A la taula es veu que sempre hi ha algu a E, pero no pretenia ser una norma general del load aware, corregeixo el paragraf perque es vegi mes clar.}

\begin{figure}[ht]
    \centering
    \includegraphics[width=0.48\textwidth]{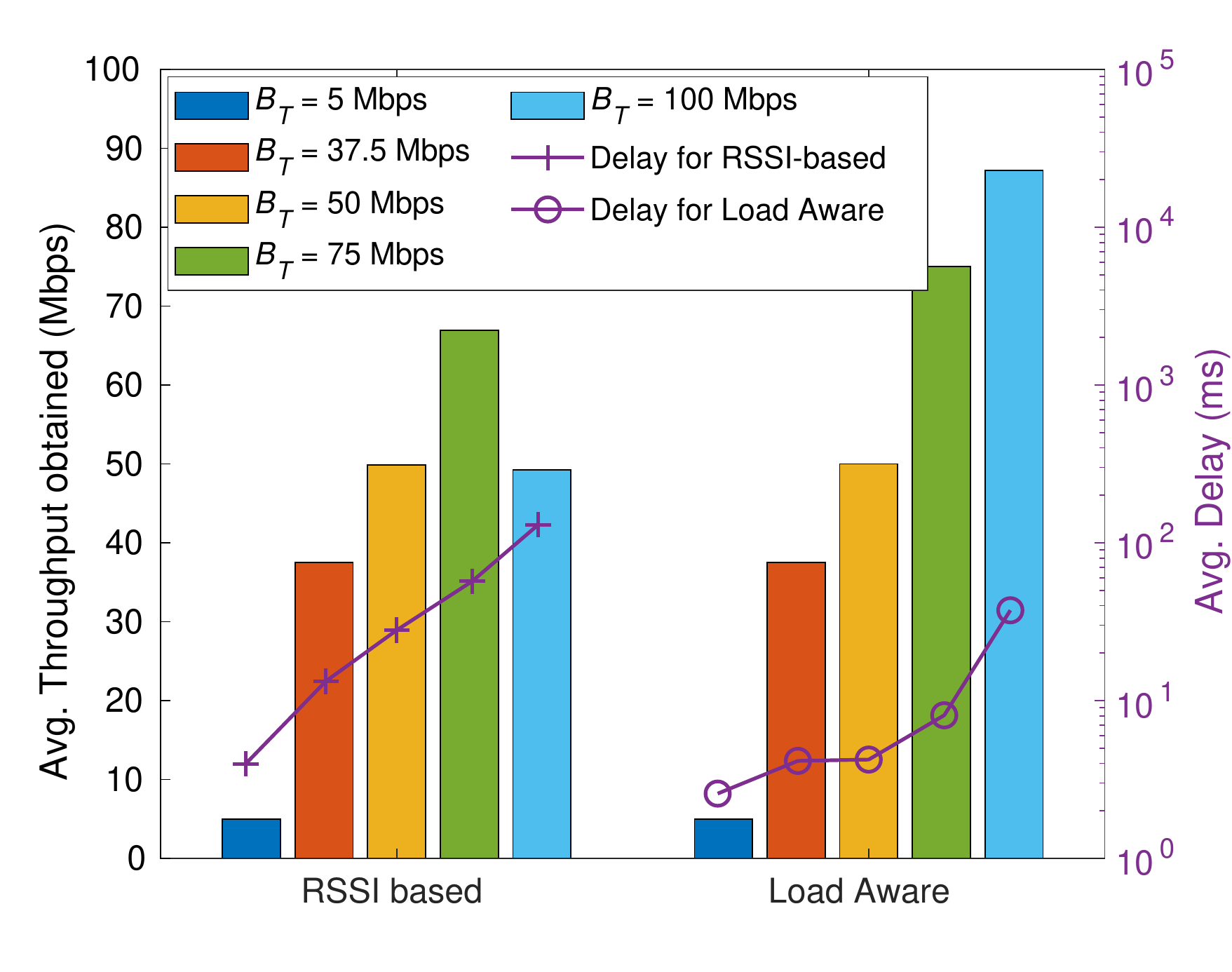}
    \caption{Throughput and delay achieved in Testbed \#2.}
    \label{s2brief}
\end{figure}

Figure \ref{s2brief} shows the results obtained for each AP/Extender selection mechanism. For $B_T = 5$ Mbps, $B_T = 37.5$ Mbps and $B_T = 50$ Mbps, both the \textit{RSSI-based} and the \textit{channel load aware} mechanisms achieved 100\% of desired throughput. However, only the \textit{channel load aware} mechanism was capable of reaching 100\% for $B_T = 75$ Mbps, with the \textit{RSSI-based} mechanism reaching only 66.9 Mbps. Finally, although the network was always congested for $B_T = 100$ Mbps, the \textit{channel load aware} mechanism managed to boost the throughput from 49.22 Mbps to 87.18 Mbps. %\textcolor{Plum}{\sout{(an increase of 77.12\%')}}.

%In terms of latency, using the \textit{RSSI-based} association mechanism always results in worse values. For 5 Mbps we obtain 3.971 ms with the RSSI-based method, while 2.573 ms are obtained with the \textit{channel load aware}, a 35.2\% decrease. For higher traffic loads the difference becomes even higher, as the \textit{channel load aware} mechanism keeps the latency below 10 ms for 5, 15, 37.5 and 50 Mbps, while the \textit{RSSI-based} method reaches 13.267 ms at 15 Mbps, and goes up to 57.026 ms for 50 Mbps. At 100 Mbps the latency is 130.24 ms and 37.345 ms (a 71.32\% decrease) for the \textit{RSSI-based} and \textit{channel load aware} mechanisms respectively.

%Nou paràgraf:
In terms of delay, the \textit{channel load aware} mechanism always had the minimum values. As a matter of example, in the worst case, with $B_T = 100$ Mbps, the delay was equal to 130.24 ms and 37.34 ms for the \textit{RSSI-based} and the \textit{channel load aware} mechanisms, respectively.

%\textcolor{blue}{(I wouldn’t go so deep on results...  just showing the general trend and the reasons of this trend.).} \textcolor{red}{As before, while I agree on this, I think it's ok in this case to provide some numbers.}

%\textcolor{Plum}{I think we can show all relevant information in section 6 with the 3 images referenced. I have left the other options for discussion. Figure \ref{s1regs} and Figure \ref{s2Regs} contain regressions of both scenarios. As they cannot be shown in a log scale (because they have negative values) I have not referenced them.  Figures \ref{s1comp}, \ref{s1APcomp} and \ref{s1avg} show the same data in different ways, I chose \ref{s1avgcomp} for the moment. If another delay figure is needed for section 6.2 I think Figure \ref{s2DelaysUnified} is a good candidate, but it clashes with Figure \ref{s1STADelay}. }

%\textcolor{blue}{Regressions are cool, but I think that we have too few points to do them... As for alternative plots, I like those showing throughput as a percentage (for instance, Figure \ref{s2Thr&DelayPerc}), but it is also true that in simulations I have used total throughput instead... so probably I should change those ones too...}

%\textcolor{blue}{(I would also include here a summary paragraph with the main insight from Testbed 2).}

%Nou paràgraf:
In this experiment, we have shown that the \textit{channel load aware} AP/Extender selection mechanism outperforms the network performance in Home WiFi scenarios of the \textit{RSSI-based} one in terms of throughput and delay. Furthermore, results also corroborate those obtained in previous simulations (compiled in Table \ref{tab:results_2_1}), in which the \textit{channel load aware} mechanism is shown to keep more deployments uncongested.  %\textcolor{Plum}{(Aquestes referencies son perilloses, perque les simulacions son amb clients repartits uniformes, i la figura 7 mostra que RSSI i channel load van casi igual amb un sol extender. El nostre testbed ni s'acosta a uniforme, i la diferencia en performance es molt mes grossa.)}

\section{The future of Home WiFi networks with multiple AP/Extenders}
\label{challenges}

In the last years, the emergence of a plethora of new applications and services in addition to the necessity of ubiquitous communication have made Home WiFi networks be more densely populated with wireless devices. Consequently, WiFi traditional spectrum at 2.4 GHz band has become scarce, and it has been necessary to extend the WiFi paradigm into new bands operating at 5 GHz and 6 GHz, with much higher resources availability.

Next generation WiFi amendments such as IEEE 802.11ax and IEEE 802.11be are taking advantage of these new bands of free license-exempt spectrum to develop physical PHY/MAC enhancements that provide Home networks with higher capacity, lower delay, and higher reliability, thus expanding WiFi into next-generation applications from the audiovisual, health care, industrial, transport, and financial sector, among others.

Nonetheless, regardless the operating band, the increasing demand of wireless resources in terms of throughput, bandwidth, and for longer connection periods makes crucial to take into consideration the interplay not only with other devices from the same Home WiFi network, but also with overlapping networks when accessing to the shared medium, including other AP/Extenders belonging to the same WLAN. In this last case, the proliferation of WLAN management platforms as discussed in Section \ref{selection} may facilitate the coordination of the network, as well as with the help of some new features coming in IEEE 802.11ax and IEEE 802.11be amendments, such as spatial reuse, OFDMA, and target wake time (TWT) solutions \cite{nurchis2019target}, including their cooperative multi-AP/Extender counterparts.

For WLANs with multiple AP/Extenders, there are still many open challenges to properly design and implement real-time \textit{load balancing} schemes among AP/Extenders when considering STA (and AP) mobility and traffic heterogeneity, including UL and DL traffic. Particularly, to create a potentially effective AP/Extender selection mechanism adapted to the aforementioned conditions, its decision metric(s) should be enriched with new parameters describing the instantaneous state of available AP/Extenders such as the number of hops to the AP, the packet latency, the available rate(s), the bit error rate (BER), or even the distance to the targeted STA.

In this last regard, the IEEE 802.11az Task Group (TGaz) aims at providing improved absolute and relative location, tracking, and positioning of STAs by using fine timing measurement (FTM) instead of signal-strength techniques \cite{wang2018specification}. Specifically, FTM protocol enables a pair of WiFi cards to estimate distance between them from round-trip timing measurement of a given transmitted signal.

%at is expected to be released in March 2020

%802.11az – Next Generation Positioning (NGP). The goal is improved absolute and relative location, tracking, and positioning of client devices (stations) using Fine Timing Measurement (FTM, based on round-trip timing measurement of a given transmitted signal) in place of the signal-strength techniques commonly used today. A final standard is expected in 2021.

%On January 2015 the Next Generation Positioning study group, IEEE 802.11az, had started its activity to address the needs of a station to identify its absolute and relative position to another station or stations it's either associated or unassociated with.

Lastly, and in line with what was stated in Section \ref{selection}, there is wide scope for the introduction of ML techniques into the AP/Extender selection mechanism. Particularly, the weight(s) of the decision metric(s) could be determined through ML, either dynamically according to a real-time observation and feedback process on the network state, or by applying the values corresponding to the most similar case from a set of predetermined patterns and scenarios.

%----------------------------------------------------------------------------------
%----------------------------------------------------------------------------------
%----------------------------------------------------------------------------------
%----------------------------------------------------------------------------------

\section{Conclusions}
\label{conclusions}

The \textit{RSSI-based} AP selection mechanism, used by default in IEEE 802.11 WLANs, only relies on the signal strength received from available APs. Therefore, in spite of its simplicity, it may result in an unbalanced load distribution between AP/Extenders and, consequently, in a degradation of the overall WLAN performance.

Though several alternatives can be found in the literature addressing this issue, the \textit{channel load aware} AP/Extender selection mechanism presented in this article stands out by its \textit{feasibility}, as it is fully based on the already existing IEEE 802.11k/v amendments, without requiring to modify the firmware of end devices to facilitate real implementation.

The potential of the \textit{channel load aware} mechanism is shown through simulations and real testbed results. It is able to outperform the traditional \textit{RSSI-based} mechanism in multi-channel scenarios consisting of multiple AP/Extenders in terms of throughput, delay, and number of situations that are satisfactorily solved, thus extending the WLAN operational range in, at least, 35\%. 

Furthermore, results from a real testbed show that the throughput is boosted up to 77.12\% with respect to the traditional \textit{RSSI-based} mechanism in the considered setup. As for the measured delay, it is consistently lower with the \textit{channel load aware} mechanism, with differences ranging from 1.398 to 92.895 ms. 

\bibliography{Bib}
\bibliographystyle{ieeetr}

\begin{IEEEbiography}[{\includegraphics[width=1in,height=1.25in,clip,keepaspectratio]{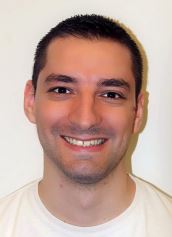}}]{Toni Adame} received his MSc degree in Telecommunications Engineering from the Universitat Politecnica de Catalunya (UPC) in 2009. He is currently a Senior Researcher in the Department of Information and Communication Technologies (DTIC) at Universitat Pompeu Fabra (UPF), responsible for the design of technical solutions in R\&D projects based on heterogeneous wireless technologies. He also collaborates as an associate lecturer in several IT degrees at UPF and Universitat Oberta de Catalunya (UOC).
\end{IEEEbiography}

\begin{IEEEbiography}[{\includegraphics[width=1in,height=1.25in,clip,keepaspectratio]{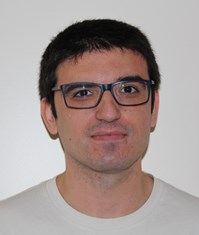}}]{Marc Carrascosa} obtained his B.Sc. degree in Telematics Engineering (2018) and a M.Sc. in Intelligent and Interactive Systems (2019) from Universitat Pompeu Fabra (UPF). He is currently a PhD student in the Wireless Networking Research Group in the Department of Information and Communication Technologies (DTIC) at UPF. His research interests are related to performance optimization in wireless networks.
\end{IEEEbiography}

\begin{IEEEbiography}[{\includegraphics[width=1in,height=1.25in,clip,keepaspectratio]{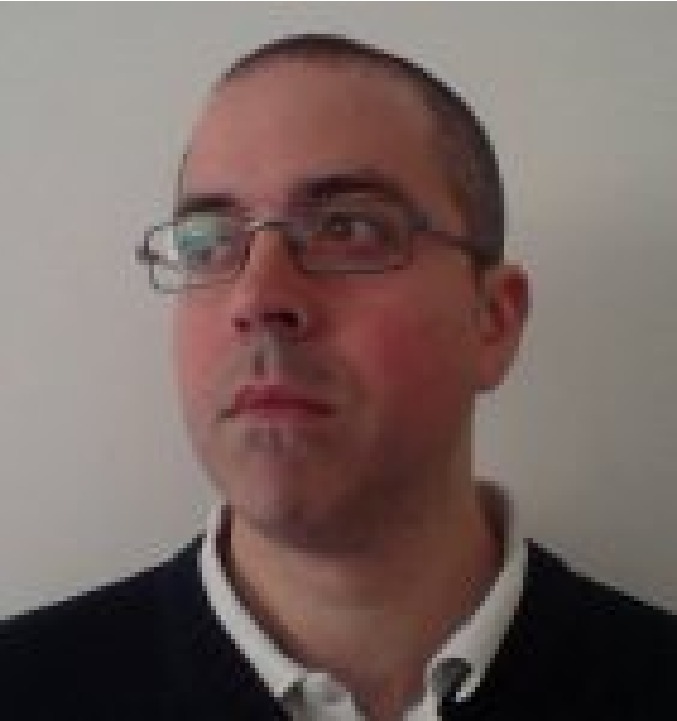}}]{Boris Bellalta} is an Associate Professor in the Department of Information and Communication Technologies (DTIC) at Universitat Pompeu Fabra (UPF). He is the head of the Wireless Networking research group at DTIC/UPF.
\end{IEEEbiography}

\begin{IEEEbiography}[{\includegraphics[width=1in,height=1.25in,clip,keepaspectratio]{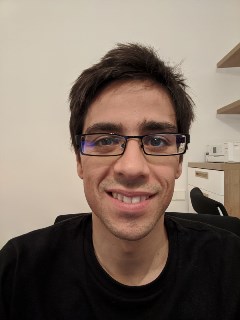}}]{Iván Pretel} is a Research Engineer in Fon Labs. He received the M.Sc in Development and Integration of Software Solutions from the University of Deusto in 2010 and the Ph.D. in Computer Engineering and Telecommunications in 2015. In 2008, he began his research career in MORElab Research Group inside the Deusto Foundation where he started as research intern in the mobile services area participating in more than 20 international and national research projects related to system architectures, Human-Computer Interaction and Societal Challenges. His research interests are in the area of data science and advanced mobile services. He is currently involved in research projects related to data science and 5G technologies, such as the 5GENESIS H2020 project. He also collaborates in several master degrees as associate lecturer at University of Deusto, giving several courses on mobile platforms, big data and business intelligence.
\end{IEEEbiography}

\begin{IEEEbiography}[{\includegraphics[width=1in,height=1.25in,clip,keepaspectratio]{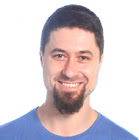}}]{Iñaki Etxebarria} is R\&D engineer at Fon Labs S.L. He obtained his degree in Telecommunications Engineering from Escuela Superior de Ingeniería de Bilbao (ETSI). He has developed his professional career in private business, before Fon he worked at Erictel M2M working on IoT, embedded equipment development and fleet management software solutions. He has been working at Fon Labs since 2015 where he has specialized in communication networks, specifically WiFi, developing innovation projects on product and technology. He has worked on several projects in international consortiums integrating WiFi in 5G networks. He currently combines engineering work with the management of the Fon Labs team.
\end{IEEEbiography}

\EOD

\end{document}